\documentclass[
superscriptaddress,
showpacs,
amsmath,
amssymb,
aps,
prb,
floatfix,
twocolumn,
]{revtex4}

\usepackage{graphicx}
\usepackage{dcolumn}
\usepackage{bm}
\usepackage{hyperref}
\usepackage{cleveref}

\begin{document}


\title{Definitive Surface Magnetotransport Study of SmB$_{6}$}

\author{Y. S. Eo}
	\thanks{These authors contributed equally.}
    \affiliation{University of Michigan, Dept.~of Physics, Ann Arbor, Michigan 48109-1040, USA}
    \affiliation{University of Maryland, Maryland Quantum Materials Center and Department of Physics, College Park, Maryland, USA}
    \email{eohyung@umd.edu, swolgast@umich.edu}

\author{S. Wolgast}
	\thanks{These authors contributed equally.}	
	\affiliation{University of Michigan, Dept.~of Physics, Ann Arbor, Michigan 48109-1040, USA}
    
\author{A. Rakoski}
	\affiliation{University of Michigan, Dept.~of Physics, Ann Arbor, Michigan 48109-1040, USA}
	
\author{D. Mihaliov}
    \affiliation{University of Michigan, Dept.~of Physics, Ann Arbor, Michigan 48109-1040, USA}
    
\author{B. Y. Kang}
	\affiliation{Gwangju Institute of Science and Technology, Department of Materials Science and Engineering, Gwangju, 61005, Republic of Korea}
    
\author{M. S. Song}
	\affiliation{Gwangju Institute of Science and Technology, Department of Materials Science and Engineering, Gwangju, 61005, Republic of Korea}
    
\author{B. K. Cho}
	\affiliation{Gwangju Institute of Science and Technology, Department of Materials Science and Engineering, Gwangju, 61005, Republic of Korea}

\author{M. Ciomaga Hatnean}
	\affiliation{Department of Physics, University of Warwick, Coventry CV4 7AL, United Kingdom}

\author{G. Balakrishnan}
\affiliation{Department of Physics, University of Warwick, Coventry CV4 7AL, United Kingdom}

\author{Z. Fisk}
	\affiliation{University of California at Irvine, Dept.~of Physics and Astronomy, Irvine, California 92697, USA}    
	
\author{S. Saha}
    \affiliation{University of Maryland, Maryland Quantum Materials Center and Department of Physics, College Park, Maryland, USA}
    
\author{X. Wang}
    \affiliation{University of Maryland, Maryland Quantum Materials Center and Department of Physics, College Park, Maryland, USA}
    
\author{J. Paglione}
    \affiliation{University of Maryland, Maryland Quantum Materials Center and Department of Physics, College Park, Maryland, USA}
    
\author{\c{C}. Kurdak}  
	\affiliation{University of Michigan, Dept.~of Physics, Ann Arbor, Michigan 48109-1040, USA}
\date{\today}

\begin{abstract}

After the theoretical prediction that SmB$_6$ is a topological Kondo insulator, there has been an explosion of studies on the SmB$_6$ surface. However, there is not yet an agreement on even the most basic quantities such as the surface carrier density and mobility. In this paper, we carefully revisit Corbino disk magnetotransport studies to find those surface transport parameters. We first show that subsurface cracks exist in the SmB$_6$ crystals, arising both from surface preparation and during the crystal growth. We provide evidence that these hidden subsurface cracks are additional conduction channels, and the large disagreement between earlier surface SmB$_6$ studies may originate from previous interpretations not taking this extra conduction path into account. We provide an update of a more reliable magnetotransport data than the previous one (Phys. Rev. \textbf{B} 92, 115110) and find that the orders-of-magnitude large disagreements in carrier density and mobility come from the surface preparation and the transport geometry rather than the intrinsic sample quality. From this magnetotransport study, we find an updated estimate of the carrier density and mobility of 2.71$\times$10$^{13}$ (1/cm$^2$) and 104.5 (cm$^{2}$/V$\cdot$sec), respectively. We compare our results with other studies of the SmB$_6$ surface. By this comparison, we provide insight into the disagreements and agreements of the previously reported angle-resolved photoemission spectroscopy, scanning tunneling microscopy, and magnetotorque quantum oscillations measurements. 
\end{abstract}


\maketitle

\section{Introduction}
 Samarium hexaboride (SmB$_6$) has attracted a great deal of interest after the theoretical prediction that it is a topological Kondo insulator. For half a century, SmB$_6$ has been known to be a Kondo insulator, the insulating gap forming by the hybridization of the 4$f$ and the 5$d$ bands. In 2010, theoretical studies predicted that some Kondo insulators, including SmB$_6$, are strong 3D topological insulators (3D TI), and therefore called topological Kondo insulators (TKI)\cite{DzeroTKI, TakimotoTKI, Dzero_2012,XiDaiPRL2013,MYe2013TKI}. If SmB$_6$ is a 3D TI, it must harbor a topologically protected surface state, or a two-dimensional metallic layer where the carriers follow a gapless Dirac-like dispersion and have a non-degenerate spin, with the direction of spin determined by the crystal momentum direction. So far, the hybridization gap and the conducting surface have been studied and verified by numerous experimental studies, including non-local transport\cite{Wolgast1}, magneto-transport\cite{Wolgast2, chineseMR}, thermal-transport, angle-resolved photoemission spectroscopy (ARPES)\cite{Jiang2ARPES2013,NXuARPES2013,NeupaneARPES2013,Denlinger2013TempARPES,Denlinger2016consistency,Xu2014direct,Frantzeskakis_ARPES}, de Haas-van Alphen (dHvA) quantum oscillations\cite{GLiDhVA, Xiang_dHvA}, point contact spectroscopy\cite{LosAlamosThermal}, scanning tunneling spectroscopy\cite{HarvardSTM, SWirthSTM}, ac conductivity\cite{NLaurita}, and spin-resolved transport\cite{song2016spin,kim2018electrical}. Compared to the weakly correlated 3D TIs, one promising aspect of SmB$_6$ from the electrical transport perspective is that the bulk is truly insulating even in the presence of mild disorder created by off-stoichiometry, and therefore its surface states can be reliably accessible without worrying about interruption by conduction through the bulk states\cite{eo2019transport}. 

  Although there has been an explosion of experimental studies on SmB$_6$ that verify the existence of conducting surface states, as mentioned above, only a few researchers have made use of this exciting material for further studies,\cite{lee2016observation,lee2018observation} perhaps because there are still remaining disagreements. Some experimental reports suggest that the surface states have a trivial origin instead of emerging from the topologically non-trivial bulk\cite{hlawenka2018samarium}. An even more serious problem is the disagreement on the very existence of the 2D metallic surface state. Arguably, the disagreement between the dHvA quantum oscillation studies is at the center of this problem. The report by Tan $et$ $al.$\cite{TanDhVa} claims that their observed quantum oscillations are from exotic 3D bulk states and that they include low-frequency ranges ($\sim$ 400 T) that are similar to the previous 2D quantum oscillation report by G. Li $et$ $al$\cite{GLiDhVA}. The two groups later follow up for further studies and the disagreement continues\cite{Xiang_dHvA, hartstein2018fermi}. A more recent study by a third group (S. Thomas $et$ $al$. \cite{thomas2019quantum}) reports that aluminum inclusions embedded in the SmB$_6$ bulk are responsible for the quantum oscillations and show a similar frequency range and 2D-like angle dependence. Next, a careful look within the reports that claim the verification of the 2D conducting surface states reveals that there are also disagreements, which may be too detailed for the community studying outside of SmB$_6$ to follow. For example, the size of the Fermi pockets estimated from the 2D dHvA reports by G. Li $et$ $al.$\cite{GLiDhVA} do not strictly agree with the size that is estimated from the ARPES reports. Also, from the historical perspective of the 2D electron gas studies, dHvA quantum oscillations have been more challenging to observe than the SdH oscillations, because the total magnetic moments are very small in a typical 2D electron layer \cite{JPEisenstein}. Although not a disagreement in the strict sense, it is peculiar that Shubnikov de-Haas (SdH) quantum oscillations through dc electrical transport have not been observed even up to $\sim$80 T,\cite{WolgastBulk} although dHvA has been reported with an onset lower than 10 T. In electrical transport, weak anti-localization (WAL), which is a supporting evidence of the helical spin structure on the TI surface, has been reported\cite{SThomasWAL,nakajima2016one,IndiaWAL}; however, there is also a report that WAL may not be universal possibility because of the magnetic impurities existing on the surface\cite{Wolgast2}. As we will show in this study, there are a number of transport experiments such as Hall effect, thermal transport, etc., but these electrical transport experiments report conductivity ($\sigma_{2D}$), carrier density ($n_{2D}$) and mobility ($\mu_{2D}$) in a wide (orders of magnitude) range of different values. 

\begin{figure}[t]
\begin{center}
\includegraphics[scale=1]{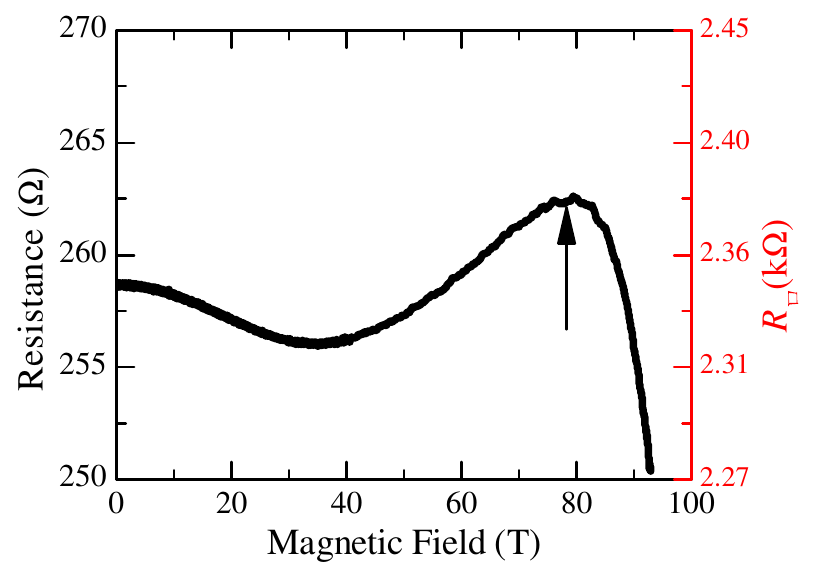}
\caption[Absence of Shubnikov de-Haas oscillations in SmB$_{6}$ up to 80 T.]{Magnetoresistance trace measured on a Corbino disk on the (001) surface using a pulsed magnetic field up to 80 T at 1.4 K. The negative magnetoresistance above 80 T, indicated with the arrow, is due to a surface-to-bulk crossover behavior, where the bulk channel starts to dominate the transport, consistent with the reduction of activation energy at high magnetic fields. \cite{WolgastBulk} The sample does not exhibit Shubnikov de-Haas oscillations up to 80 T. The corresponding sheet resistance (below 80 T) is shown in the right scale (in red).
}
\label{Fig:NoSdH93T}
\end{center}
\end{figure}

 The first possibility for such disagreements is that the quality varies from sample to sample. Of course, one can naturally think that a different crystal growth method can introduce different types of defects and disorder. This aspect must be resolved especially for the quantum oscillation perspective. The dHvA reported from G. Li $et$ $al$.\cite{GLiDhVA} was measured on a sample that was grown by the aluminum flux method, whereas the report from Tan $et$ $al$.\cite{TanDhVa} was measured on a sample that was grown by the optical floating zone method. Another aspect is the role of magnetic impurities. A common belief is that if SmB$_{6}$ is a true 3D TI, protected by time-reversal symmetry, magnetic impurities and the surface state existence should have an antagonistic relation to each other. However, it is not clear if the existence of the magnetic impurities in the crystal is the most dominant factor for influencing the surface properties. In addition, even within the samples grown by the same technique, there is a wide perception that the qualities of the bulk and surface may differ from one to another. Another possibility is that the data was interpreted differently. So far, when the surface carrier density and the mobility are estimated by transport experiments on SmB$_6$, they rely on the classical (Drude) or semi-classical (Boltzmann) transport models. For example, when using these classical or semi-classical models to interpret transport experiments, 
 the geometric conversion factor ($g$) from measured resistance to conductivity ($\sigma_{2D}$) or resistivity ($\rho_{2D}$), must be known accurately. SmB$_6$ is difficult in a sense that many transport measurements were performed on thick (order of tens of microns up to mm) single crystal samples. Thus, unlike the weakly correlated 3D TIs that are grown in thin films, the side surfaces, edges, and corners on the sample surface also contribute as current paths, and the interpretation of $g$ can be very difficult. The problem becomes even more difficult when studying magnetotransport on the conducting surface, where the angle between the current and the magnetic field direction is different for all surfaces. Also, unlike other 2D systems, electron and hole pockets can simultaneously exist in the surface Brillouin zone in a 3D TI as long as the number of Fermi pockets is odd. If electron and holes coexist, the total carrier density can be dramatically overestimated by the Hall coefficient. The strategy to avoid this complication is using a Corbino disk, which confines the surface current path and is insensitive to the carrier species. There are two reports which measured surface transport using Corbino disks, and each estimates the $n_{2D}$ and $\mu_{2D}$ by gating\cite{PSyersCorbino} and magnetotransport\cite{Wolgast2}.
 Many early experimental studies must be re-investigated because these aspects were not taken into account at the time. In the end, the parameters ($\sigma_{2D}$, $n_{2D}$, and $\mu_{2D}$) from transport must agree with the other experimental measurements (dHvA, ARPES, and optical ac conductivity). In this paper, we first present an alarming study that suggests that subsurface cracks conduct in parallel with the exposed surfaces at low temperatures of SmB$_6$. This implies that rough polishing, which is a typical intermediate procedure for obtaining a smooth surface on a single crystal, can create cracks just below the surface, thus changing $g$. Similarly, we suggest that it is even possible that small grain boundaries on a natural as-grown crystalline surface can be unexpected conducting paths and can complicate the surface geometric factor even more. Furthermore, keeping these new aspects in mind, we perform Corbino disk magnetotransport studies on multiple samples, including samples grown by both aluminum flux and floating zone techniques, both polished and unpolished. Our results show that the sample-to-sample variation, including both flux and floating zone growth samples, is smaller than the perceived widespread disagreement. In our studies, the estimated carrier density and mobility only vary within an order of magnitude when only considering the samples that are prepared properly by our own standards. 
 Many groups have previously tried to characterize the surface conduction on SmB$_6$ using high-field magnetoresistance measurements with the hope of observing SdH oscillations. On the (001) surface, as we reported previously, we do not observe SdH oscillations up to 80 T, as shown in Fig.~\ref{Fig:NoSdH93T}. The absence of SdH oscillations by such measurements provide an upper bound for the mobility of surface electrons. Our goal in this paper is to construct an allowed parameter space, $n_{2D}$ vs. $\mu_{2D}$, for the (001) surface and estimate a range of where each pocket should exist. Furthermore, we compare this transport analysis with other experimental reports.

%

\begin{figure}[ht]
\begin{center}
\includegraphics[scale=1.2]{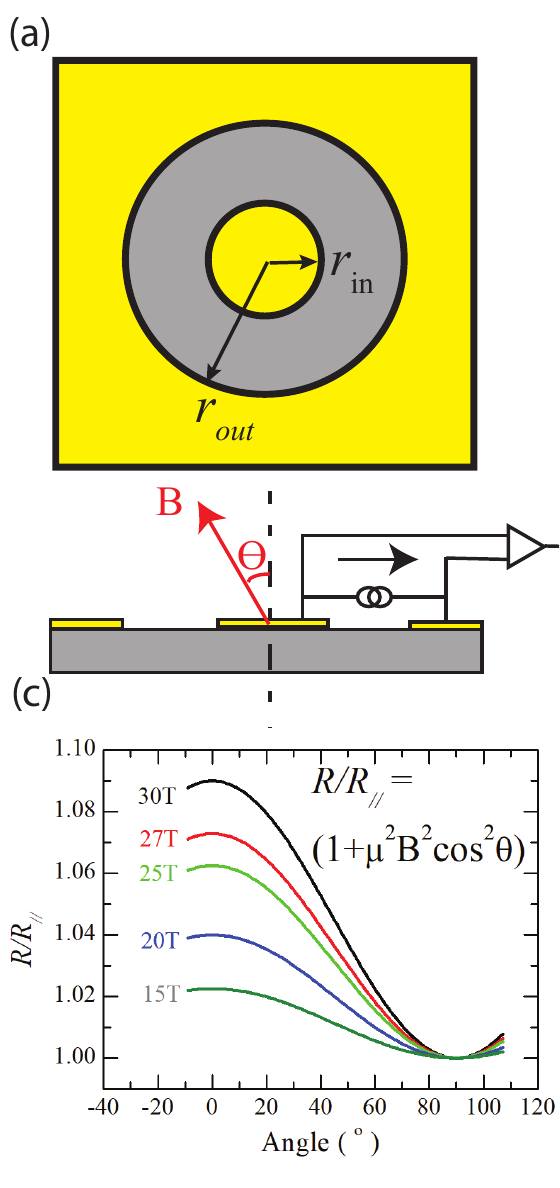}
\caption{A Corbino disk. (a) Setup of a Corbino disk on sample.  (b) Simulated resistance ratio vs. Magnetic field angle fixed at different magnetic field values based on Eq.~\ref{Eq:CorbinoFormulaBField}. }
\label{fig:Corbinosimulation}
\end{center}
\end{figure} 

	\label{Eq:linearBoltzmann3}

\section{\label{Sec:cracks}Influence of Subsurface Cracks on Surface Transport}

Within the SmB$_{6}$ community, there is a widespread perception that the surface sheet resistance estimated from the resistance plateau below 3\textendash 4 K varies dramatically from sample to sample. Of course, depending on the method of crystal growth, the leading order of defects can be different from one sample to another, and some samples can even be non-stoichiometric or contain unintentional impurities. However, we have been long suspicious of whether the measured resistance is indeed the true sheet resistance of the SmB$_{6}$ surface, especially in situations where the reported sheet resistances are very low (a few Ohms), for two reasons. First, such a low sheet resistance would imply a high electron mobility, in excess of 10,000 (cm$^2$/V$\cdot$s), which cannot be achieved unless the surface electrons are formed in an environment that is relatively free of disorder. We note that unlike semiconductor based two-dimensional electron gas (2DEG) systems, where the 2D electrons live in a clean interface of an epitaxially grown semiconductor heterostructure interface, the surface states of SmB$_6$ are exposed in ambient conditions. We note that although the surface of a 3D TI prohibits back scattering, this does not guarantee a high mobility. Absence of back scattering only results in a mild correction to the scattering time compared to the case where back scattering is allowed\cite{ozturk2017influence}. In the case of a exposed surfaces, the carriers are expected to scatter strongly from disorder arising from the non-epitaxial native oxide that must be present on all SmB$_6$ transport samples. Second, transport samples with such electron mobilities (in excess of 10,000 (cm$^2$/V$\cdot$s)) are expected to exhibit SdH oscillations starting in magnetic fields as low as 1 T, which has not been seen in any SmB$_6$ studies. We also note that this low resistance cannot be a bulk conducting path, unlike in weakly correlated 3D TIs. First, the surface of SmB$_{6}$ can be tuned by gating\cite{PSyersCorbino}. Also, from inverted resistance measurements, the bulk resistivity continues to rise exponentially below 3\textendash 4 K, and therefore we can safely conclude that the carriers flowing in the bulk are negligible\cite{eo2019transport}. Instead, we hypothesize that invisible (or hidden) conductive surfaces can exist in addition to the surfaces that are exposed. If these hidden conduction paths are not accounted for in transport experiments, this may explain why some of the experiments report low sheet resistances. 
We rely heavily on the Corbino disk structures, as shown in Fig~\ref{fig:Corbinosimulation}~(a), which allow us to study the transport properties of different crystalline surfaces individually. The resistance of the Corbino sample is expected to be inversely proportional to the sheet conductivity: 
\begin{equation}
	R_{\mathrm{Corbino}}=\frac{\ln(r_{\mathrm{out}}/r_{\mathrm{in}})}{2\pi}\frac{1}{\sigma_{s}(0)},
	\label{Eq:CorbinoFormulaBField}
\end{equation}
where $\sigma_{s}(0)$ (in 1/$\Omega$) is the surface conductivity at zero magnetic field and $r_{in}$ and $r_{out}$. In an ideal 2D layer, where the carrier density is independent of magnetic field, the Corbino resistance would be magnetic field dependent as a result of mobility reduction, following the Lorentz force factor:
\begin{equation}
	R_{\mathrm{Corbino}}(B)=\frac{\ln(r_{\mathrm{out}}/r_{\mathrm{in}})}{2\pi}\frac{(1+(\mu B \mathrm{cos} \theta)^2)}{ne\mu},
	\label{Eq:CorbinoFormulaBField2}
\end{equation}
where $B$ is the applied magnetic field, $\theta$ is the angle between the field and the surface normal direction, $n$ is the surface carrier density, $e$ is the electron charge, and $\mu$ is the surface mobility. We will later rely on these two equations above to analyze our magnetotransport data. For instructive purposes, we illustrate in Fig.~\ref{fig:Corbinosimulation}~(b) how the resistance changes with angle at a fixed magnetic field based on Eq.~(\ref{Eq:CorbinoFormulaBField}).  

In this section, we study the role of surface preparation. First, we study the role of polishing and surface treatment by oxygen plasma induced oxidization. Among these studies, the most surprising finding is that the sample surfaces that were prepared with rougher polishing tend to result in lower resistance plateau values than the ones that were more finely polished. In typical materials, we would expect the opposite. If the surface is rough, the surface roughness can contribute to extra scattering. Because this scattering will contribute to lowering the mobility, the resistance is expected to be higher. 
 
We hypothesized in this study that subsurface cracks that are created by rough polishing can serve as additional surface conduction paths. If SmB$_{6}$ is truly a 3D TI, then since subsurface cracks are a termination of the bulk, they must also be topologically protected surfaces. This hypothesis is consistent with the resistance plateau value trend that we observed for different polishing qualities because, in semiconductors, it is well known that rougher polishing creates subsurface cracks with larger length scales\cite{PeiCrack,mchedlidze1995subsurface}. 
 
\begin{figure}[ht]
\begin{center}
\includegraphics[scale=0.9]{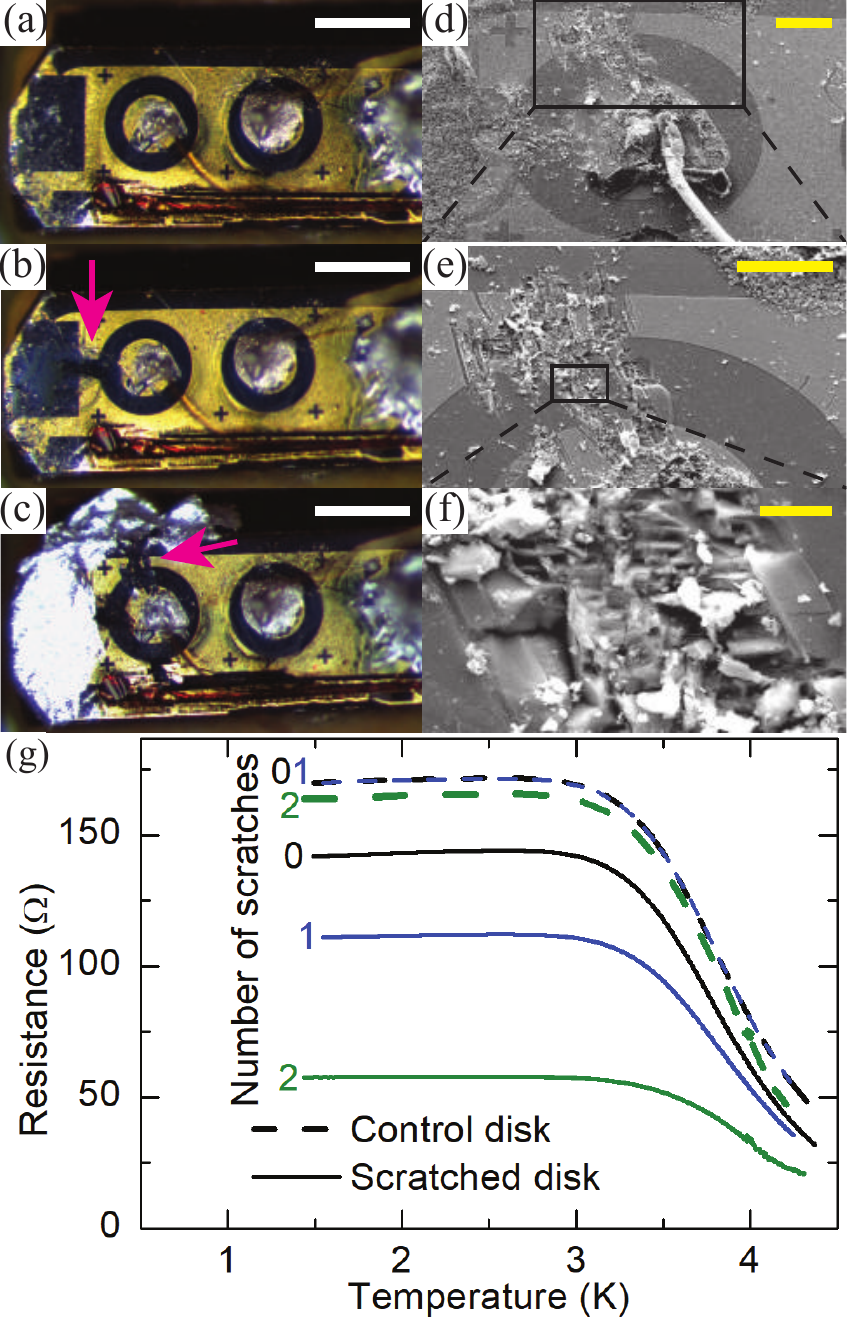}
\caption[Scratched Corbino measurements.]{(Color) Scratched Corbino measurements. (a) -- (c) Optical images of the two Corbino disks before scratching and after each scratch. The scratches are indicated with magenta arrows. White scale bars are 500 $\mu$m. (d) -- (f) SEM images of the second scratch at increasing magnification. Yellow scale bars are 100 $\mu$m, 100 $\mu$m, and 10 $\mu$m, respectively. (g) Resistance vs. temperature curves of both disks before scratching and after each scratch. Thick solid lines indicate the scratched disk, while thin dotted lines indicate the unscratched (control) Corbino disk. (Sample: Fisk011)} 
\label{Fig:Crack}
\end{center}
\end{figure}

To test the hypothesis that subsurface cracks contribute to surface conduction, we prepared an SmB$_{6}$ sample with two Corbino disks fabricated on a finely polished surface. The resistance vs$.$ temperature was measured below $\sim$4.5 K, both before and after the active region (annular ring) of one of the Corbino disks was scratched with a scriber, as shown in Fig.~(\ref{Fig:Crack})~(a) $-$ (c). As shown in Fig.~(\ref{Fig:Crack})~(g), while the plateau resistance of the unscratched Corbino disk remains almost identical in both measurements within a few percent, the plateau resistance of the scratched Corbino disk further lowered from 140 $\mathrm{\Omega}$ to 110 $\mathrm{\Omega}$. After the second scratch, the resistance further lowers to 60 $\mathrm{\Omega}$. Indeed, the surface roughness increases after scratching, but the resistance does not become higher. Instead, it becomes lower, consistent with our hypothesis that the rougher surface conducts. 
	
\begin{figure}[ht]
\begin{center}
\includegraphics[scale=0.9]{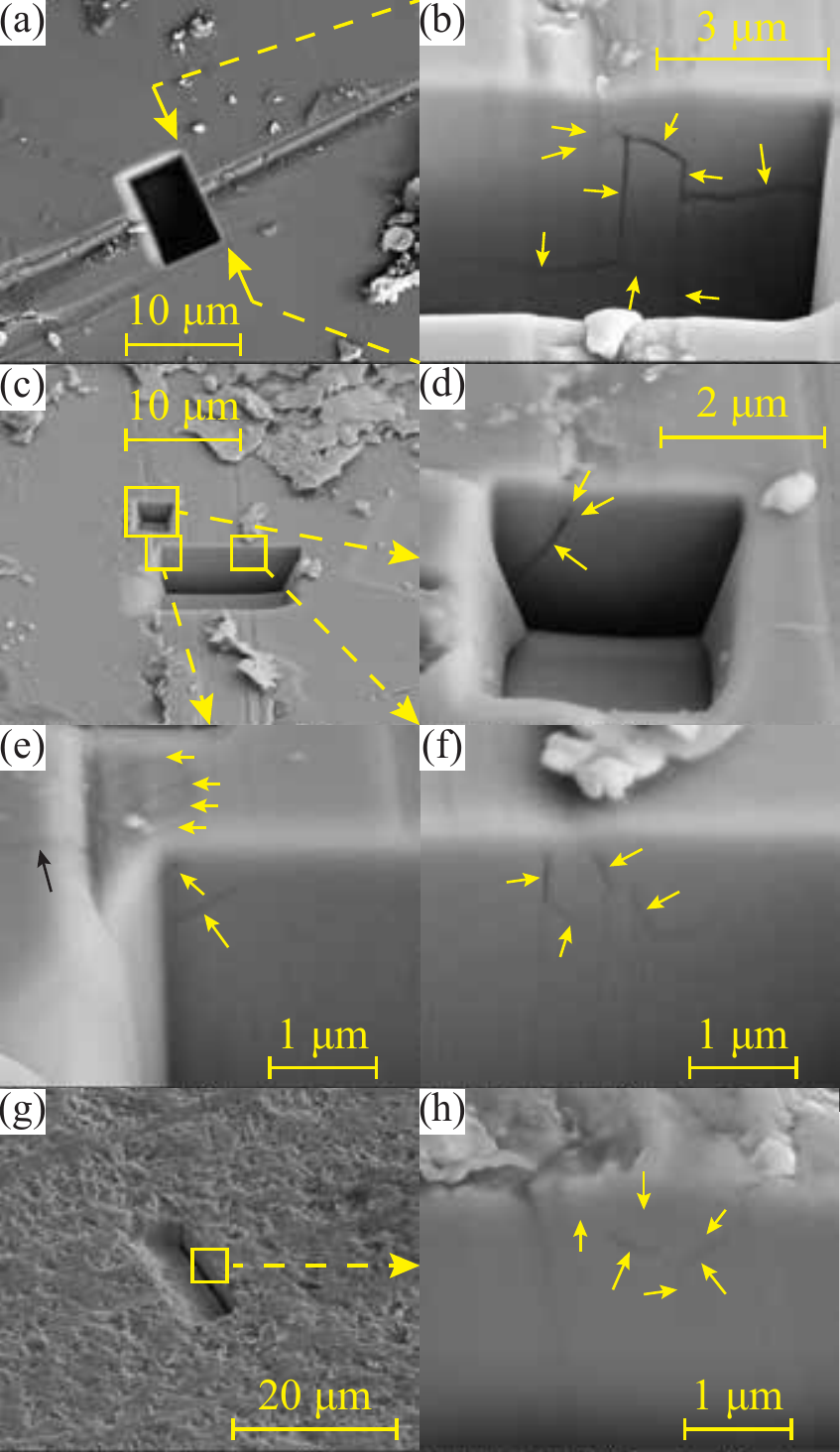}
\caption{(Color) Ion-milled cross-sections. (a) Ion-milled cross-section across a scratch. (b), Subsurface cracks visible below the scratch in (a). (c) Ion-milled cross-sections across another scratch. (d) - (f), Subsurface cracks visible below the scratch in (c). (g), Ion-milled cross-section on a rough-polished surface. (h), Subsurface cracks visible below the rough surface in (g). In all panels, small yellow arrows indicate cracks. (Sample: Fisk011)} 
\label{fig:Cracks}
\end{center}
\end{figure}

We further investigated if the scratched surface indeed harbors subsurface cracks. An ion-beam milling was performed on the scratched surface. The ion-milled wall profile is shown in Fig.~(\ref{Fig:Crack})~(d) $-$ (f) through an SEM image. We indeed observed subsurface cracks that are several microns long and up to 100 nm wide. Normally, we would expect that the polishing grit particles would introduce a stress to the surface, and therefore create subsurface cracks. Indeed, subsurface cracks were also observed on a separate SmB$_{6}$ crystal prepared by rough-polishing (P1200 grit, which produces micron level surface roughness), as shown in Fig.~(\ref{fig:Cracks})~(g) - (h). The cracks visible in Fig.~(\ref{fig:Cracks})~(h) are up to 1-$\mu$m long in the transverse or vertical direction, though they are much narrower than those seen in Fig.(~\ref{fig:Cracks}) (b) and (d), approaching the focal resolution limit of our SEM. Although cracks produced at finer polishing levels are below the resolution of our SEM images, it is not unreasonable to hypothesize their existence and contribution to the total surface conduction. Such cracks are expected to form whenever the maximum contact stress, which actually does occur a few $\mu$m below the surface,\cite{hertz1882beruhrung} exceeds the tensile strength of the material.
    
The length scales of these subsurface cracks would scale with the size of the grit particles. Therefore, the sample must be polished with the finest possible grit size and thinned sufficiently to eliminate the subsurface cracks that are created from the rougher polishing grit introduced in the previous polishing step. We eventually chose a grit size of 0.3 $\mu$m of Al$_{2}$O$_{3}$ for the finest polishing step. 

\begin{figure}[ht]
\begin{center}
\includegraphics[scale=0.9]{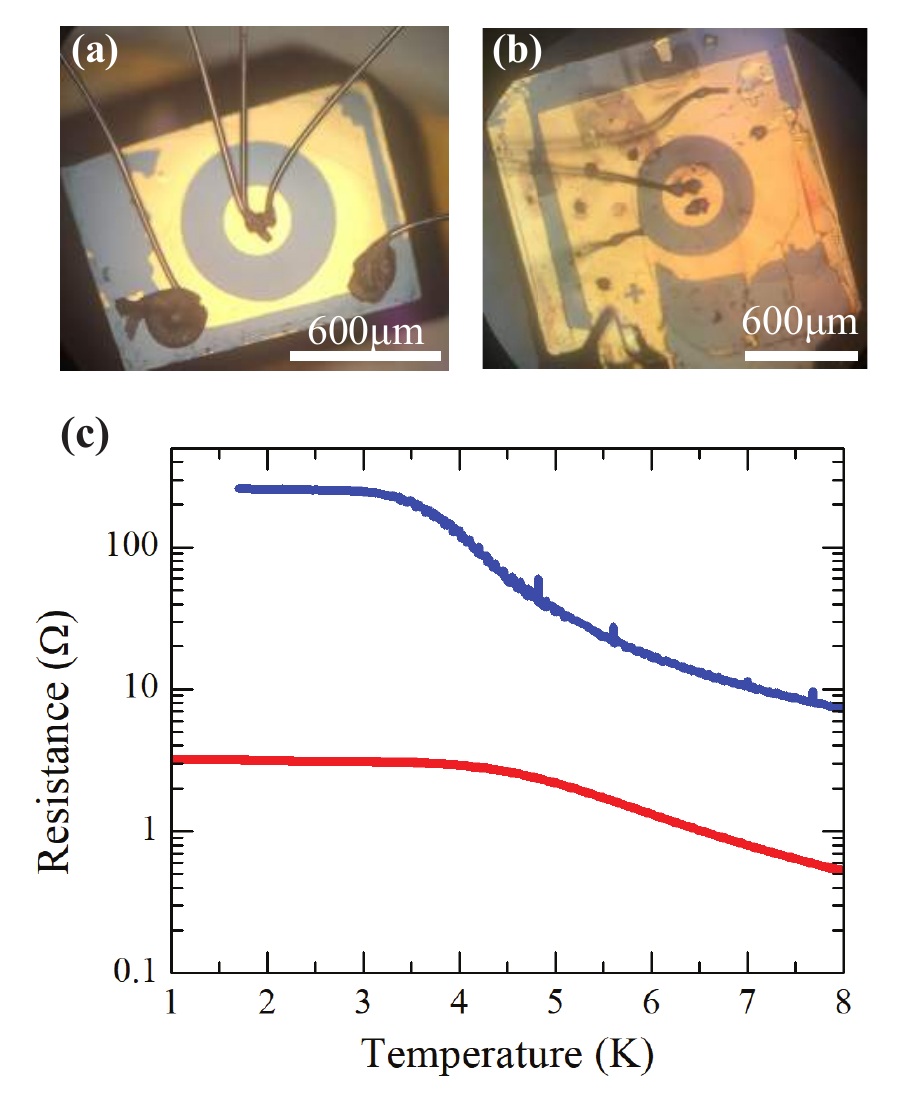}
\caption[Comparison of resistance vs$.$ temperature of a Corbino disk on unpolished single crystal vs$.$ polycrystal.]{Comparison of resistance vs$.$ temperature of a Corbino disk on unpolished single crystal vs$.$ polycrystal. (a) Corbino disk on a clean single crystal surface in the (001) direction. (b) Corbino disk on a clean polycrystalline surface with many domain boundaries. (c) Resistance vs$.$ temperature of the two samples. The blue curve is measured from sample (a), and the red curve is measured from sample (b).(Sample: (a) GISTasGrown2 (b) GISTasGrown1)} \label{Fig:AsGrown}
\end{center}
\end{figure}

Next, we consider as-grown surfaces. If thinning and polishing can potentially create cracks, a single crystal that begins with a large enough surface is desirable. We were fortunate to obtain a single crystal that had a large enough area to fabricate a Corbino disk, as shown in Fig.~(\ref{Fig:AsGrown})~(a). In comparison, we also fabricated a Corbino disk on a low quality sample where multiple crystal facets can be seen with boundary lines between them (shown in Fig.~(\ref{Fig:AsGrown})~(b)). These lines are most likely naturally grown cracks or domain boundaries. Neither samples underwent any thinning or polishing treatments. The samples were only cleaned in dilute HCl before and after the lithography process. We measured the resistance vs$.$ temperature on both samples using the standard Corbino disk geometry. The results are shown in Fig.~(\ref{Fig:AsGrown})~(c). The resistance curve of the single crystalline surface is shown in the blue line, and the result for the polycrystalline surface is shown in the red line. The single crystal shows a resistance plateau of $\sim$260 $\mathrm{\Omega}$, with a corresponding sheet resistance of 2.3 k$\mathrm{\Omega}$. This sheet resistance value is consistent with the results from the final polishing step of 0.3 micron Al$_{2}$O$_{3}$. This gives us confidence that the final polishing step was sufficient for our studies. On the other hand, the resistance plateau value of the polycrystalline sample is only 3.2 $\mathrm{\Omega}$. We believe that the domain boundaries serve as conduction paths, similar to that of the conduction through subsurface cracks. 

Another aspect to consider is the role of disorder on the surface. Previous studies report that a native oxide, most likely Sm$_2$O$_3$, forms on the surface of SmB$_6$\cite{PRL_Tjeng}. This oxide is suspected to be a source of surface magnetism, as reported previously\cite{nakajima2016one,Wolgast2}. As an instructive exercise, we treat our Corbino disk sample with a series of plasma oxidization events to introduce native oxides with more disorder and acid etching to remove the native oxide layer. 

\begin{figure}[ht]
\begin{center}
\includegraphics[scale=1.2]{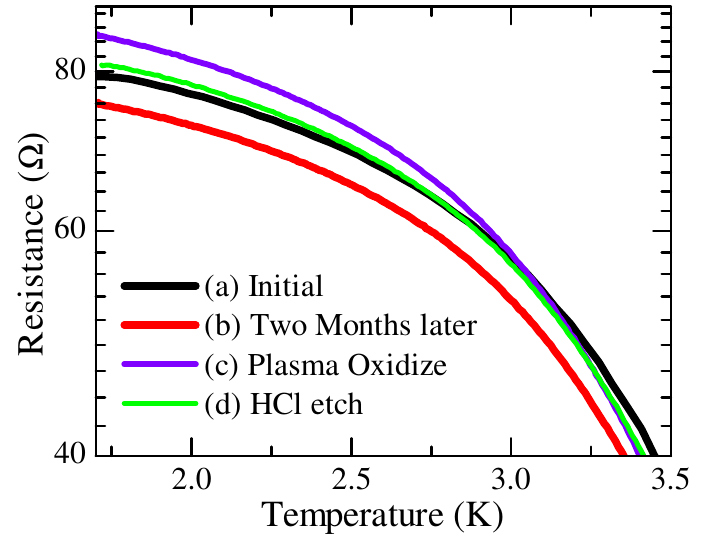}
\caption[Surface oxidization experiment]{Surface oxidization experiment on a Corbino disk on SmB$_6$.(Sample:Paglione79B)} 
\label{Fig:Oxide}
\end{center}
\end{figure}
    
To study this effect, we prepared a Corbino disk patterned on a polished SmB$_6$ sample. We summarize the results shown in Fig.~(\ref{Fig:Oxide}). First, after etching the sample in HCl, we measured resistance vs. temperature within an hour after the etching event, shown in curve Fig.~(\ref{Fig:Oxide})~(a). The resistance value rises to a maximum value of 79 $\Omega$ at base temperature. After exposing the sample in air for two months, the resistance measurement was repeated, as shown in curve Fig.~(\ref{Fig:Oxide})~(b). Then, the maximum value of resistance changed to 75 $\Omega$. We again measured the sample after plasma oxidizing with 100W power under 0.5 torr of oxygen for 10 minutes, and measured the resistance of the sample again shown in curve Fig.~(\ref{Fig:Oxide})~(c). The resistance rises up to a maximum value of 85 $\Omega$. Finally, we etched this sample with HCl diluted down to 20$\%$ for 4 minutes. We loaded this sample into the cryostat within 40 minutes and remeasured resistance vs. temperature, shown in curve Fig.~(\ref{Fig:Oxide})~(d). The maximum resistance this time was 81 $\Omega$. The sample does show a small change with plasma treatment and acid etching. However, we find these effects are much smaller than the previous scratch experiments.

\begin{figure}[ht]
\begin{center}
\includegraphics[scale=0.7]{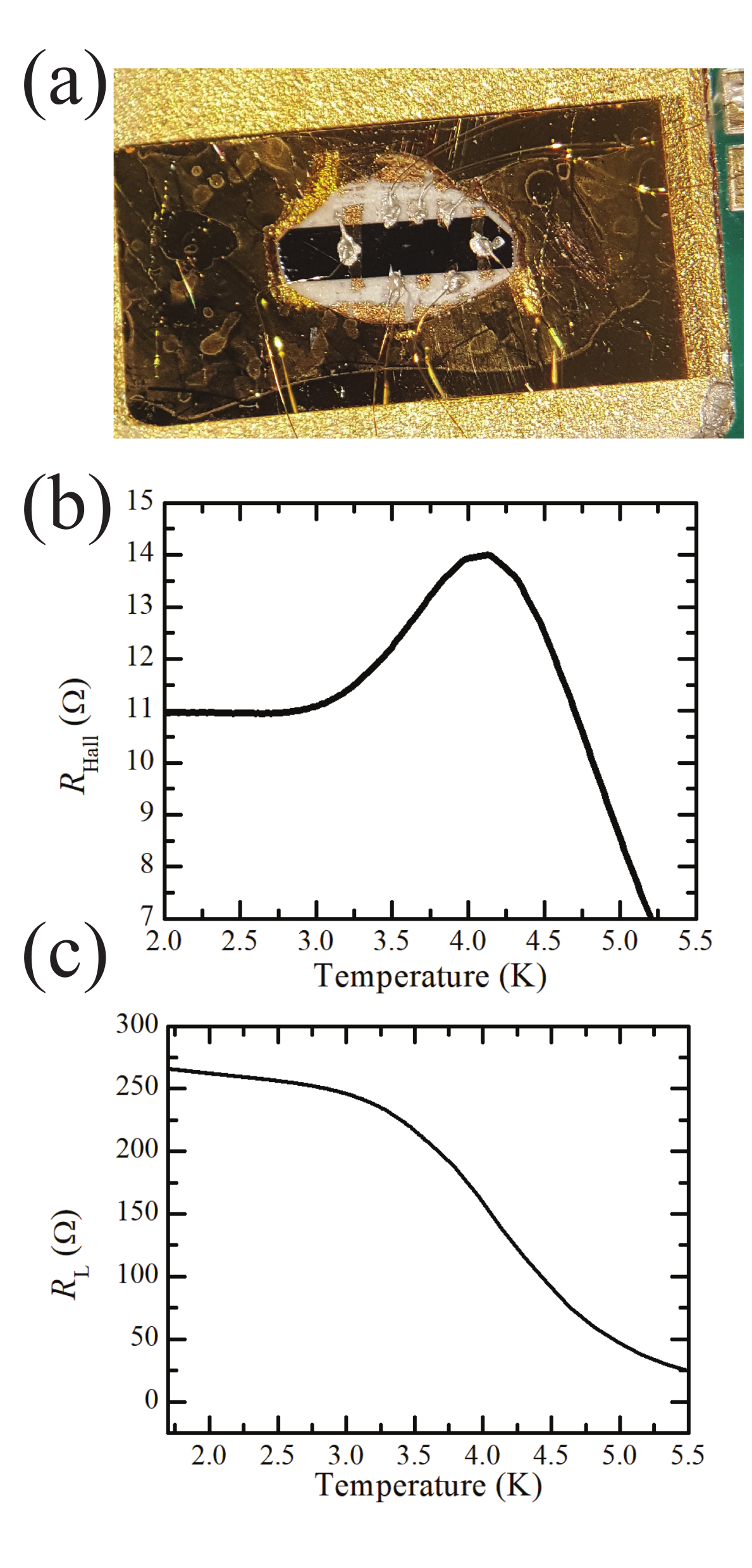}
\caption[Hall Measurement]{Hall bar measurement of SmB$_6$ at 14T. (a) Hall bar sample. (b) Hall resistance at 14T vs. temperature. (c) lateral resistance vs. temperature. (Sample: GISTHall1)}
\label{Fig:AlexaHall}
\end{center}
\end{figure}

We next consider how surface preparation and transport geometry can influence magnetotransport of surface studies of SmB$_6$. First, we discuss samples with a Hall bar geometry. In our earlier reports, we have attempted to find the surface carrier density using a bulk Hall bar geometry\cite{Wolgast1}. Surprisingly, the surface carrier density estimation by a naive Hall voltage formula ($V_{H}=BI/n_{2D}e$) results in an extremely large range of $n_{2D} \sim 10^{18} (\mathrm{cm}^{-2})$. We will later show in Sec.~\ref{Sec:ParameterSpace} that this is an unphysical value. As an instructive demonstration, we prepared a sample more carefully, polishing four surfaces as identically as possible with the goal of eliminating the subsurface cracks. This sample is shown in Fig.~\ref{Fig:AlexaHall}, and the measurement resulted in a more reasonable carrier density of $n_{2D} = 1.7\times 10^{14}(\mathrm{cm}^{-2})$. However, we find that many errors can be introduced in interpreting the data. We are relying on the assumption that the sample is a perfect Hall bar with four identical surface qualities, ignoring the current flowing through the side surfaces and the edges, and the contribution of contact size. 
    
To correctly find the geometric factor ($g$) of the transport geometries, we must know where and how much the current flows through each surface. Choosing a Corbino disk geometry confines the current path to flow on a single surface and therefore we can ignore the surfaces where current does not flow. In our previous study\cite{Wolgast2}, we have found that a Corbino disk on a SmB$_6$ surface does not follow $R \propto (1+\mu^2B^2cos^2\theta)$ as in Eq.~\ref{Eq:CorbinoFormulaBField} because the sheet resistance without the Lorentz force contribution has an overall negative magnetoresistance, independent of its tilt angle, i.e., $dR_\square (|B|)/dB <0$. Still, resistance divided by the in-plane magnetoresistance ($\theta = 90^{0}$), $R_{||}$, shows a good agreement with $R/R_{||} \propto (1+\mu^2B^2cos^2\theta)$. 
	
\begin{figure}[ht]
\begin{center}
\includegraphics[scale=0.45]{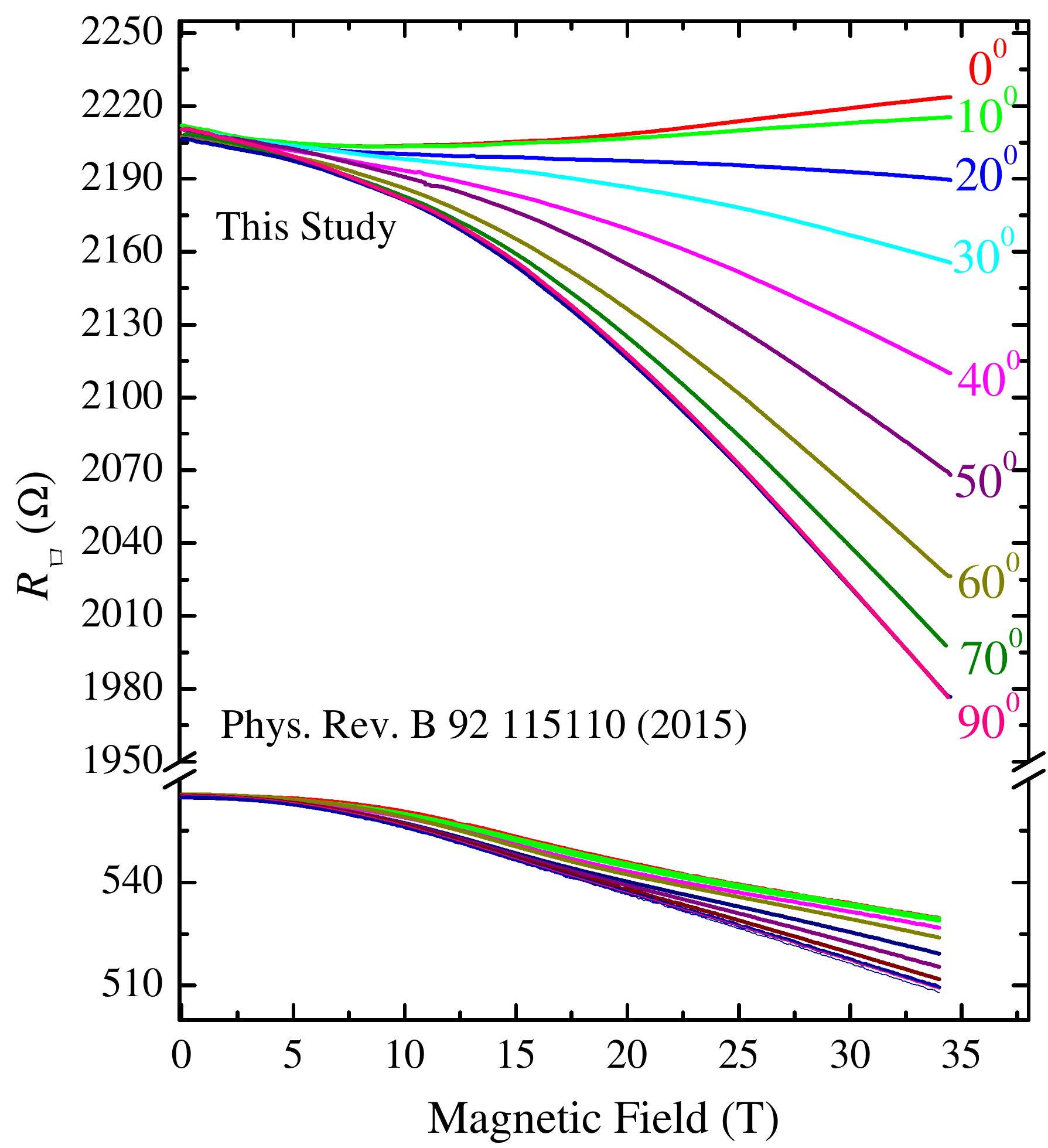}
\caption[Comparing magnetotransport of SmB$_{6}$ with previous work]{Comparing magnetotransport of Corbino disk on a (001) surface of SmB$_{6}$ from our previous report\cite{Wolgast2} and our current result. The temperature is at 0.35 K, and the magnetic field is swept while the angle of the magnetic field is kept fixed. (Sample: FiskGate1)} 
\label{Fig:MRBeforeAfter}
\end{center}
\end{figure}

When the sample is finely polished, we find that the $R/R_{||} \propto (1+\mu^2B^2cos^2\theta)$ effect becomes larger. We study how our previously reported Corbino disk SmB$_6$ sample\cite{Wolgast2} changes its magnetotransport characteristics after carefully polishing. In our previous work\cite{Wolgast2}, when we were not aware of subsurface crack conduction, the Corbino disk sample was prepared by final polishing with a SiC polishing pad with an average grit size of 2.5 $\mathrm{\mu}$m\cite{Wolgast1}. In our present work, the Corbino disk sample was prepared by final polishing with an Al$_{2}$O$_{3}$ oxide with a particle size of 0.3 $\mathrm{\mu}$m. It is also important that the finer polishing step after the rough polishing step must remove the subsurface cracks created from the previous rough polishing step. Our Corbino disk magnetotransport measurements on the (001) surface at 0.35K before and after the finer polishing is shown in Fig.~\ref{Fig:MRBeforeAfter}. We first find that the estimated sheet resistance is much larger after fine polishing. Also, while the negative magnetoresistance remains -10$\%$ up to 35T, estimated by the in-plane magnetic field sweep, the angle dependence is more dramaticafter fine polishing.
    
\begin{figure}[ht]
\begin{center}
\includegraphics[scale=0.8]{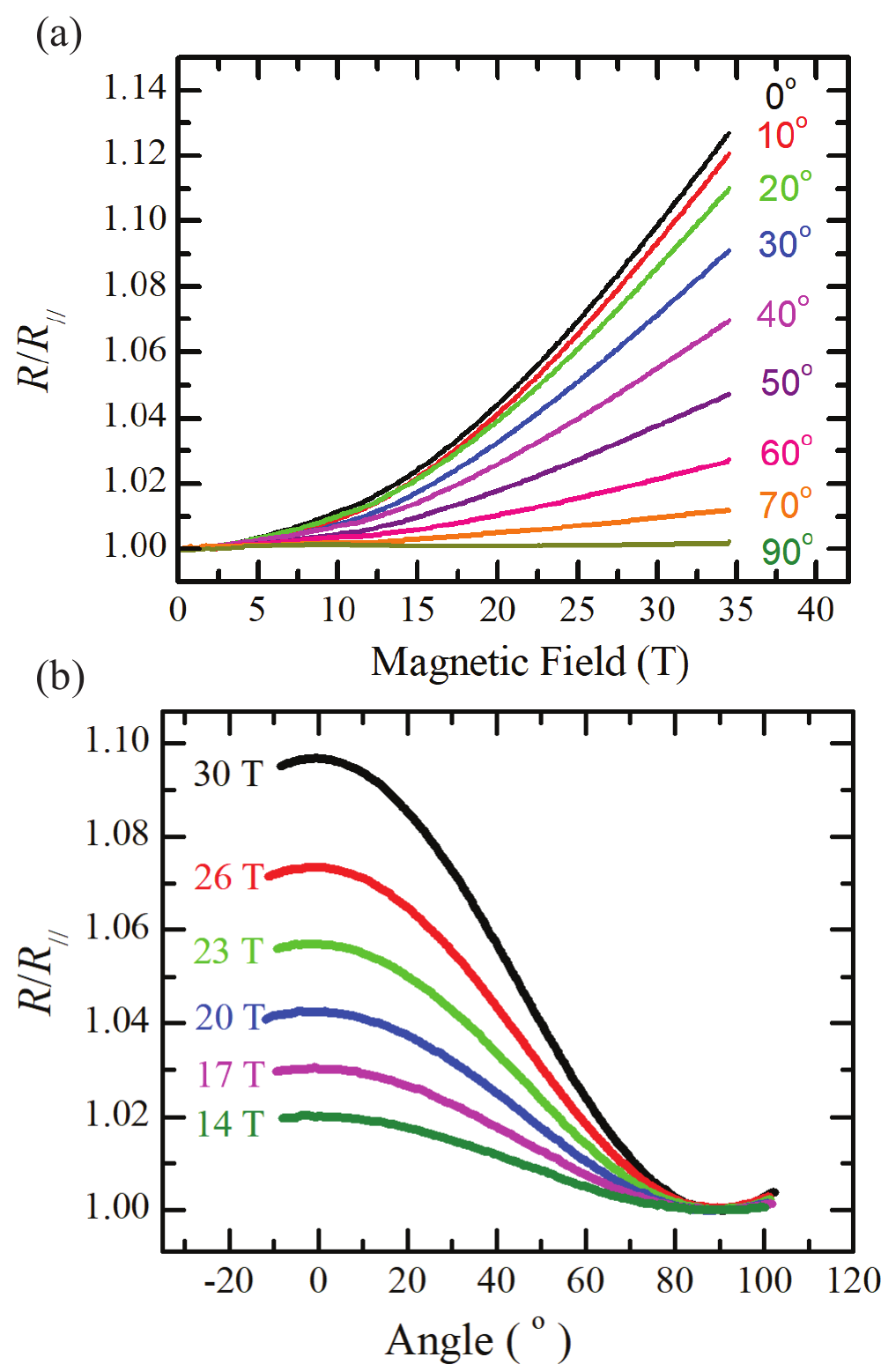}
 \caption[Corbino Magnetotransport of SmB$_{6}$]{Magnetotransport results on a Corbino disk on the (001) surface. The temperature is at 0.35 K. (a) Dividing the resistance vs. magnetic field from the top curves in Fig.~\ref{Fig:MRBeforeAfter} to resistance at 90$^0$ (in-plane). (c) Resistance vs. tilt angle of the magnetic field at different magnetic field values. (Sample: FiskGate1)}
\label{Fig:MR001Results}
\end{center}
\end{figure}

Next, we show that this enhanced angle-dependent magnetotransport after surface re-preparation changes the estimation of the surface mobility and the carrier density.  As shown in Fig.~(\ref{Fig:MR001Results})~(a), the trace for the out-of-plane direction (0$^{\circ}$ angle, shown in a black line) deviates from a quadratic behavior. Also, the in-plane direction (90$^{\circ}$ angle, shown in a green line) shows a negative magnetoresistance in contrast to Eq.~\ref{Eq:CorbinoFormulaBField}, which predicts that there should be no magnetoresistance at that angle ($\theta$ = 90$^{0}$). In our previous report, we have speculated that the carrier density increases as the bulk band gap narrows\cite{Wolgast2} and that the mobility changes by the Kondo effect and surface roughness scattering \cite{Wolgast2}. As mentioned above, in spite of the presence of the negative magnetoresistance, the classical two-dimensional magnetotransport behavior is still valid. When dividing the resistances by the in-plane magnetic field resistance ($R(\theta)/R_{||}$), the ratio shows a good agreement with the quadratic dependence, as shown in Fig.~(\ref{Fig:MR001Results})~(b). When rotating the angle of the sample while the magnetic field is fixed, the data follows the cosine squared behavior, as shown in Fig.~(\ref{Fig:MR001Results})~(c). Fig.~(\ref{Fig:MR001Results})~(b) and Fig.~(\ref{Fig:MR001Results})~(c) also show that the surface carriers still experience the Lorentz force on a two dimensional layer. From Fig.~(\ref{Fig:MR001Results})~(a), this implies that the surface is an unusual 2D system in that the carrier density and mobility change as the magnetic field increases, also consistent with our previous report\cite{Wolgast2}. A quadratic fit to Fig.~(\ref{Fig:MR001Results})~(b) results in a mobility of 104.5 (cm$^{2}$/V$\cdot$sec) and a carrier density of 2.71$\times$10$^{13}$ (1/cm$^2$). From cosine square fits of Fig.~(\ref{Fig:MR001Results})~(c), the carrier density changes by about 10$\%$, and the mobility changes by about 3$\%$ over the range of magnetic fields, consistent with our previous studies \cite{Wolgast2}.
Lastly, we summarize in Table~(\ref{Tab:TransportEvolution}) how our transport parameter estimations have evolved from Hall bar geometry to Corbino disk geometry, and then further through carefully preparing the surface. Comparing to our previous report with this sample from Ref.~\cite{Wolgast2}, as shown in Fig.~(\ref{Fig:MRBeforeAfter}), the estimated surface mobility increased by 41$\%$ and the carrier density decreased by a factor of 7.4. Although our most recent mobility estimation is higher than what we had originally estimated, this value is still small compared to high-quality 2DEG systems. We believe this is why SdH oscillations have not yet been observed. In the following section, we discuss our most recent magnetotransport results. Just by relying on classical magnetotransport analysis, we did not observe clear signatures of the existence of multiple channels in magnetotransport. This is in contrast to other experimental reports such as ARPES and dHvA that report on multiple pockets, so we will also discuss why our results do not show signatures of multiple channels based on the low mobility of the surface carriers.
\begin{table}[ht]
\centering 
\begin{tabular}{c c c c c} 
\hline\hline 
Year& Geometry& $R_{\square}$ ($\mathrm{\Omega}$)& $n_{2D}$ (cm$^{-2}$)& $\mu_{2D}$ (cm$^{2}$/V$\cdot$sec) \\ [1ex]
\hline 
2012 & Hall bar\cite{Wolgast1}&9.1 & 1.0$\times$10$^{18}$ & 0.69\\ 
2015 & Corbino Disk\cite{Wolgast2} &570 & 2.0$\times$10$^{14}$ & 61\\
Current & Hall bar  &1400 &1.7$\times$10$^{14}$ & 26\\
Current & Corbino Disk  &2200 & 2.71$\times$10$^{13}$ & 105\\[1ex] 
\hline 
\end{tabular}
\caption[Evolution of the surface transport parameter estimation from measurements at different years.]{Surface transport parameter estimation from measurements after understanding effects of geometry and subcrack conduction.}
\label{Tab:TransportEvolution}
\end{table}
\section{\label{Sec:ParameterSpace}Constructing the Transport Parameter Space of Each Channel}
\begin{figure}[t]
\begin{center}
\includegraphics{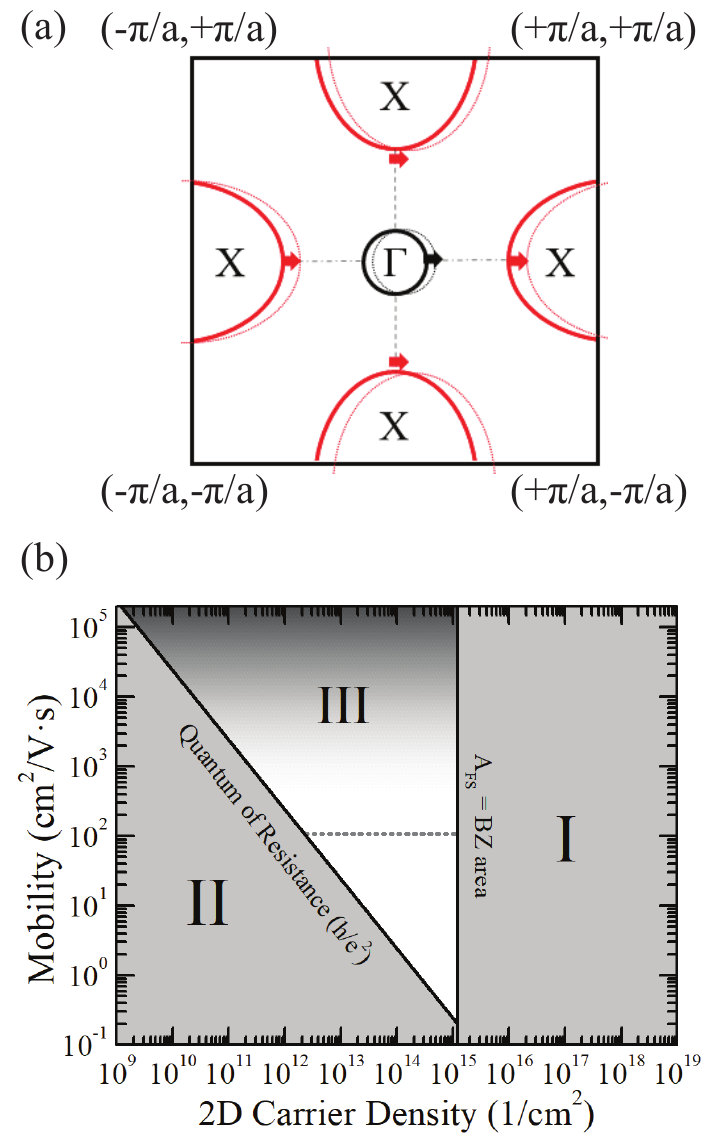}
\caption[(001) surface of SmB$_{6}$.]{(001) surface of SmB$_{6}$. (a) The three Fermi pockets of the BZ on the (001) surface. The solid red lines indicate the $X$-pockets, and the solid black line indicates the $\Gamma$-pocket. The dotted lines indicate how the Fermi surfaces deform if the electric field is applied in the $k_x$-direction. (b) The surface transport parameter space of SmB$_{6}$ in the (001) direction. The gray area (region I and II) are the forbidden regions where the parameters cannot exist. The dotted line indicates the mobility of 1/(80 T). Region III is shown in a gradient color, with the darker area indicating that the transport parameters would be less likely to exist at higher mobilities.}
\label{Fig:FermiPocketAndFundamentalLimit}
\end{center}
\end{figure}

The topological Kondo insulator theory predicts that three surface Fermi pockets exist, each surrounding a high symmetry point in the Brillouin zone (BZ)\cite{XiDaiPRL2013,MYe2013TKI}. In particular, on the (001) crystal surface, theory predicts that one of the Fermi pockets should be centered at the $\Gamma$-point (called the $\Gamma$-pocket in short), and the other two should be centered at the $X$-point (or $X$-pocket in short), as shown in solid red in Fig.~(\ref{Fig:FermiPocketAndFundamentalLimit})~(a). If SdH oscillations had been observed, the verification of each pocket would have been possible. As mentioned previously, SdH oscillations were not observed up to 80 T, and therefore our analysis is based on classical magnetotransport. The quadratic fit of $R(B)/R_{||}$ at a fixed angle and a cosine squared fit of $R(\theta)/R_{||}$ at a fixed magnetic field relying on Eq.~(\ref{Eq:CorbinoFormulaBField}) assumes that only a single channel (or Fermi pocket) exists in the BZ. 

In our magnetotransport results shown in Fig.~(\ref{Fig:MR001Results}), the reason we were not able to observe distinct signatures of multiple channels can be explained if the mobilities of each channel are small. If an electric field is applied in the $+k_x$ direction in the BZ, the Fermi surfaces will change as shown in the dotted lines in Fig.~(\ref{Fig:FermiPocketAndFundamentalLimit})~(a). To simplify, we consider that the top and bottom $X$-pockets contribute equally to the left and right $X$-pockets by assuming that the associated scattering time does not depend on the direction of crystal momentum. Also, we assume that intrapocket scattering is dominant compared to interpocket scattering. Then, the conductivity of the two channels, $\sigma_{\Gamma}$ and $\sigma_{2X}$, add together to get the total conductivity ($\sigma_{t}$): 
 \begin{equation}
 \sigma_{t} = \sigma_{\Gamma} + \sigma_{2X}.
 \label{Eq:totalcondCh3}
 \end{equation}
Furthermore, when a magnetic field is applied, we assume the carrier density and mobility that we measure are effective parameters ($n_{\mathrm{eff}}$ and $\mu_{\mathrm{eff}}$) based on both pockets, i.e., $\sigma_{t}=n_{\mathrm{eff}}e\mu_{\mathrm{eff}}$. The magnetoconductivity can also be expressed in terms of the transport parameters of the two individual pockets:
\begin{equation}
 \frac{n_{\mathrm{eff}}e\mu_{\mathrm{eff}}}{1+\mu_{\mathrm{eff}}^2B^2\cos^2\theta}
 =\frac{n_{\Gamma}e\mu_{\Gamma}}{1+\mu_{\Gamma}^2B^2\cos^2\theta}
 +\frac{n_{2X}e\mu_{2X}}{1+\mu_{2X}^2B^2\cos^2\theta},
\label{Eq:totalmagncondCh3}
\end{equation}
where $n_{\Gamma}$ ($n_{2X}$) is the surface carrier density of the $\Gamma$-pocket (two $X$-pockets), and $\mu_{\Gamma}$ ($\mu_{2X}$) is the mobility of the $\Gamma$-pocket (two $X$-pockets). $R(B)/R_{\mathbin{\!/\mkern-5mu/\!}}$ at a fixed angle or $R(\theta)/R_{\mathbin{\!/\mkern-5mu/\!}}$ at a fixed magnetic field would be inversely proportional to Eq.~(\ref{Eq:totalmagncondCh3}). The inverse of Eq.~(\ref{Eq:totalmagncondCh3}) can be expanded in a series, and dividing by $R_{\mathbin{\!/\mkern-5mu/\!}}$ ($=1/\sigma_t$) we have
\begin{equation}
\begin{split}
\frac{R(B,\theta)}{R_{\mathbin{\!/\mkern-5mu/\!}}}
&= 1 + \mu_{\mathrm{eff}}^2B^2\cos^2\theta \\
&+ (\frac{\sigma_{\Gamma}}{\sigma_{t}}-(\frac{\sigma_{\Gamma}}{\sigma_{t}})^2)(\mu_{\Gamma}^2-\mu_{2X}^2)^2B^4\cos^4\theta +\dotsb,
\label{Eq:EffectCondSeries}
\end{split}
\end{equation}
where
\begin{equation}
\mu_{\mathrm{eff}}=\sqrt{(\mu_{\Gamma}^2-\mu_{2X}^2)(\frac{\sigma_{\Gamma}}{\sigma_{t}})+\mu_{2X}^2}.
\label{Eq:EffectMobility}
\end{equation}
From Eq.~(\ref{Eq:EffectCondSeries}), we can see that as long as any of the mobilities ($\mu_{\Gamma}$ and $\mu_{2X}$) from the channels are much smaller than $1/B$, the third term or higher order terms in Eq.~(\ref{Eq:EffectCondSeries}) will be overwhelmed by the second order term, and Eq.~(\ref{Eq:EffectCondSeries}) will be indistinguishable from the magnetotransport for the single channel equation, Eq.~(\ref{Eq:CorbinoFormulaBField}). Therefore, we will not be able to tell if there are two channels from magnetotransport. Furthermore, the effective carrier density is: 
\begin{equation}
n_{\mathrm{eff}}=\frac{\sigma_t}{e\mu_{\mathrm{eff}}}=\frac{(n_{\Gamma}\mu_{\Gamma} + n_{2X}\mu_{2X})^{3/2}}{(n_{\Gamma}\mu_{\Gamma}^3 + n_{2X}\mu_{2X}^3)^{1/2}}.
\label{Eq:EffectCarrierDensity}
\end{equation}

We have so far explained why our magnetotransport of a Corbino disk fails to resolve signatures of multiple pockets. In the following subsections, we will show that the transport parameters from each pocket can only exist in a constrained space. Furthermore, together with our (effective) surface carrier density and mobility results, we will further constrain the range (or area) of the parameter spaces, where each pocket should exist. This result will be compared with other experimental reports. 

\subsection{Forbidden Parameter Space}

In a 2D transport, there are two fundamental limits that constrain the parameter space. First, the size of the Fermi pocket cannot be larger than the defined surface Brillouin zone of the material. When a periodic boundary condition (Born-von Karman) is chosen to describe the system of interest in quantum mechanics, there is a relation between the occupied area (or volume) in $k$-space and the associated density of carriers. In a two-dimensional system, this relation is:
 \begin{equation}
 n_{2D}=\frac{s}{(2\pi)^2}A_{\mathrm{FS}},
 \label{Eq:FermiSurfaceCarrierDensity} 
 \end{equation}
where $A_{\mathrm{FS}}$ is the area of the Fermi surface, and $s$ is the spin degeneracy ($s$ = 2 for typical materials, $s$ = 1 when the spin degeneracy is split). The maximum allowed surface carrier density is then, of course, related to the total area of the BZ. The maximum carrier density allowed on the (001) surface of SmB$_6$ is $n_{\mathrm{max}}$ = 5.86$\times$10$^{14}$ cm$^{-2}$ when there is no spin degeneracy (in a true 3D TI case). Therefore, any carrier density larger than this value is forbidden. In Fig.~(\ref{Fig:FermiPocketAndFundamentalLimit})~(b), the surface transport parameter space, where we plot the mobility and surface carrier density in a log-log graph, this forbidden region indicated as region I, shaded in gray.  
 
Next, in 2D transport, the system undergoes a metal-to-insulator transition when $R_{\square}$ becomes larger than the quantum of resistance, $h/e^2$ (Ioffe-Regel criterion for 2D). Therefore, $R_{\square}$ larger than the quantum of resistance must be forbidden. The diagonal line in Fig.~(\ref{Fig:FermiPocketAndFundamentalLimit})~(b) represents $R_{\square} = h/e^2$ (or quantum conductivity: $\sigma_s = e^2/h$), and region II, the gray shaded region that is below the diagonal line is forbidden. 
 
The absence of SdH oscillation up to 80 T, as shown in Fig.~(\ref{Fig:NoSdH93T}), also provides a constraint on the pockets in the parameter space. We first revisit the Lifshitz-Kosevich (LK) formalism very briefly. The conductivity oscillates with a damping factor that is related to the scattering of the carriers. This is the Dingle damping factor ($D_{D}$), and it is related to the mobility. The Dingle damping factor is:
\begin{equation}
D_{D}=\exp(-\frac{\pi}{\mu_Q B}),
\label{Eq:DingleDamp}
\end{equation} 
where $\mu_Q=e\tau_Q/m^*$ is the quantum mobility. In the semi-classical approach, $\tau$ is estimated by Fermi's Golden Rule where the scattering potential is multiplied by an extra factor, ($1-\cos\theta$), so that it does not include the forward scattering\cite{GrapheneScattTime,GrapheneScatteringTheory}. This factor is not included in the $\tau_Q$ (quantum scattering time) in quantum oscillations and can differ from $\tau$ found from our classical (or semiclassical) transport approach. We will revisit the LK formalism when comparing our transport to the dHvA results\cite{GLiDhVA}.
 
From the Dingle factor ($D_{D}$), we see that a low mobility means that the onset of quantum oscillations would be at a high magnetic field. Of course, the exact onset magnetic field estimation would depend on the sensitivity of the measurement, how we model the broadening of the Landau levels due to disorder, and the prefactor of the quantum scattering time, etc. A crude estimate of the onset of quantum oscillations is when $\mu B\approx 1$, or $B\approx 1/\mu$. Therefore, the absence of SdH up to 80 T, which is the highest magnetic field ever applied on the SmB$_6$ (001) surface, tells us that the magnetic field requirement for the onset of SdH oscillations must be greater than 80 T, and therefore the mobility should be in the vicinity of 1/(80 T) or less. In Fig.~(\ref{Fig:FermiPocketAndFundamentalLimit})~(b), the mobility of 1/(80 T) is shown as a horizontal dotted line, and the area above is indicated as region III. The transport parameters would be less likely to exist at a higher mobility, although this is not strictly forbidden. To graphically show this in the figure, region III is shown in a gradient color that becomes darker at higher mobility ranges.
 
Excluding the regions of I and II, and since the high mobility far above 1/(80 T) is less likely, the Fermi pockets must be in or near the white triangular region in Fig.~(\ref{Fig:FermiPocketAndFundamentalLimit})~(b). Next, continuing with the effective mobility and carrier density found from our Corbino disk magnetotransport, we constrain the parameter space of carrier density and mobility where each pocket can be allowed. 
 
\subsection{Constraining the Transport Parameter Space Region for Each Pocket from Corbino Magnetotransport}

\begin{figure}[t]
\begin{center}
\includegraphics[scale=1.1]{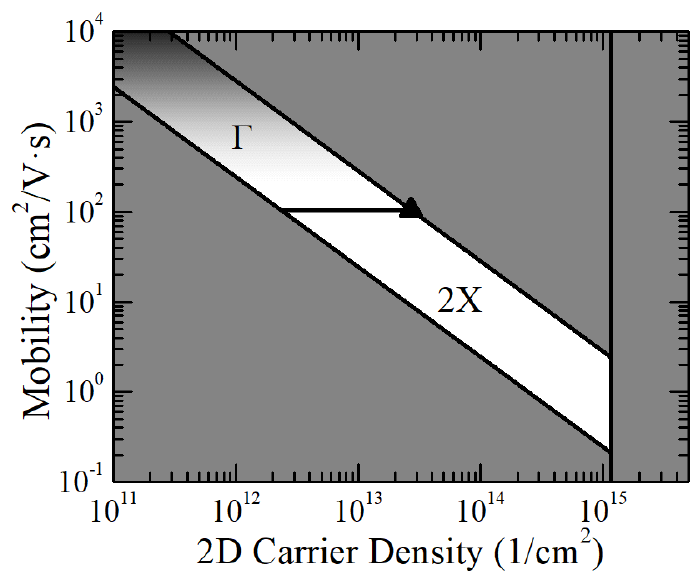}
\caption[The surface transport parameter space of SmB$_6$ on the (001) surface.]{The surface transport parameter space of SmB$_6$ on the (001) surface.}
\label{Fig:CorneredParameterSpace}
\end{center}
\end{figure}

In this subsection, we construct a parameter space region where each pocket ($\Gamma$ and $2X$) on the (001) surface can be located based on our magnetotransport results. In the transport parameter space, the carrier density and mobility of each pocket is represented as a point. Our result for effective surface carrier density and mobility ($n_{\mathrm{eff}}$ = 2.71$\times$10$^{13}$ cm$^{-2}$ and $\mu_{\mathrm{eff}}$ = 104.5 cm$^2$/(V$\cdot$s)) from the Corbino disk is shown as a black triangle in Fig.~(\ref{Fig:CorneredParameterSpace}). If we assume that our magnetotransport is a result of two channels, we can use Eq.~(\ref{Eq:totalcondCh3}). 
The corresponding conductivities of each pair of points must add up to the conductivity corresponding to the triangular point (which denotes the effective parameters). The two points that represent the transport of each channel are somewhere in the allowed region shown in Fig.~(\ref{Fig:FermiPocketAndFundamentalLimit})~(b). Also, to be consistent with Eq.~(\ref{Eq:totalcondCh3}) \textendash ~Eq.~(\ref{Eq:EffectCarrierDensity}), one of the mobilities must be larger than $\mu_{\mathrm{eff}}$ while the other mobility must be smaller (a detailed justification is in Appendix A). Then, the allowed parameter space of the channel with the small carrier density ($\Gamma$-pocket), and the channel with the large carrier density ($2X$-pocket) should be in the regions that are indicated in Fig.~(\ref{Fig:CorneredParameterSpace}). Although we cannot specify the exact carrier density and mobility values, the small carrier density channel must be in region $\Gamma$, and the large carrier density channel must be in region $2X$. In the following section, we compare our constructed parameter space to other experimental reports such as Hall, dHvA, and ARPES.
 
\section{Comparison with Other Experiments}
In this section, we compare our constructed transport parameter space from the previous section to other experimental reports on the SmB$_{6}$ (001) surface. Before we proceed, we will briefly review how to estimate the carrier density and mobility from non-transport experiments.

\subsection{Quantum Oscillation (dHvA) Interpretation }
When the area of the Fermi surface in the BZ has a size of $A_{\mathrm{FS}}$, the associated frequency of quantum oscillations, $F$, follows the Onsager relation:
\begin{equation}
F(T)=\frac{\hbar}{2\pi e}A_{\mathrm{FS}}.
\label{Eq:OnsagerCh3}
\end{equation}
Also, since the carrier density ($n$) and $A_{\mathrm{FS}}$ are related (Eq.~(\ref{Eq:FermiSurfaceCarrierDensity})), the relation between $n$ and $F$ can be found as:
\begin{equation}
n_{2D}=s\frac{e}{h}F(T),
\label{Eq:CarrierDensityQuantumOscill}
\end{equation}

According to the LK formula, the amplitude of the oscillations is damped by two factors, the temperature damping factor ($D_{T}$) and the Dingle damping factor ($D_{D}$), which was introduced previously (Eq.~(\ref{Eq:DingleDamp})). The two damping factors are given by:
\begin{equation}
D_{T}=\frac{2\pi^2 (k_{B}T/\hbar \omega_{c})}{\sinh(2\pi^2 (k_{B}T/\hbar \omega_{c}))},
\label{Eq:TempDampFactor}
\end{equation}
\begin{equation}
D_{D}=\exp(-\frac{\pi}{\mu_Q B}),
\end{equation}
where $\omega_{c}$ (=$eB/m^*$) is the cyclotron frequency. From the temperature dependence of the amplitude of quantum oscillations, the effective mass, $m^*$, can be found. From the magnetic field dependence of the amplitude at a fixed temperature, the Dingle damping factor ($D_{D}$) can be used to find $\mu_{Q}$. 

\subsection{ARPES and STM Quasiparticle Interference Interpretation}
In ARPES and STM quasiparticle interference (QPI), the momentum and energy resolved intensity data ($I(k,\omega)$) is measured and is understood by the spectral density function, $S(k,\omega)$, associated with removal of electrons from photons, with energy  $\omega$\cite{ARPESCuprates,ColemanBook}: 
\begin{equation}
I(k,\omega)=f(E)S(k,\omega),
\label{Eq:ARPESIntensity}
\end{equation}
where $f(E)$ is the Fermi-Dirac distribution and $S(k,\omega)$ is given by:
\begin{equation}
S(k,\omega)=-\frac{1}{\pi}\frac{\mathrm{Im}(\Sigma)}{[\omega - E(k) - \mathrm{Re}(\Sigma)]^2 + \mathrm{Im}(\Sigma)^2}.
\label{Eq:ARPESPSD}
\end{equation}
where $\Sigma$ is the self energy. Notice from Eq.~(\ref{Eq:ARPESIntensity}) that $I(k,\omega)$ becomes weak above the Fermi energy because of $f(E)$. Also notice that when the photon energy $\omega$ approaches $E(k)- \mathrm{Re}(\Sigma)$, the magnitude of $I(k,\omega)$ is enhanced. Therefore, ARPES can measure the energy dispersion below the Fermi energy and slightly above at finite temperatures, with the momentum dependence. By resolving different momentum directions at the Fermi energy, $E_{F}(k)$, the sizes and shapes of the Fermi pockets can be found. The size of the Fermi pockets can be converted to carrier density using Eq.~(\ref{Eq:FermiSurfaceCarrierDensity}). From the slope or curvature of the dispersion $E(k)$ below the Fermi energy, the effective mass ($m^*$) can be found. Furthermore, the spectral broadening is related to the momentum relaxation, and the associated scattering time, $\tau_p$, is:
\begin{equation}
\frac{\hbar}{\tau_{p}} = - \mathrm{Im}(\Sigma).
\label{Eq:ARPEStau}
\end{equation}
Therefore, the mobility can be found by $\mu = e\tau_{p}/m^*$.

\subsection{Estimation of Transport Parameters from Previous Reports}

\begin{figure}[t]
\begin{center}
\includegraphics[scale=1]{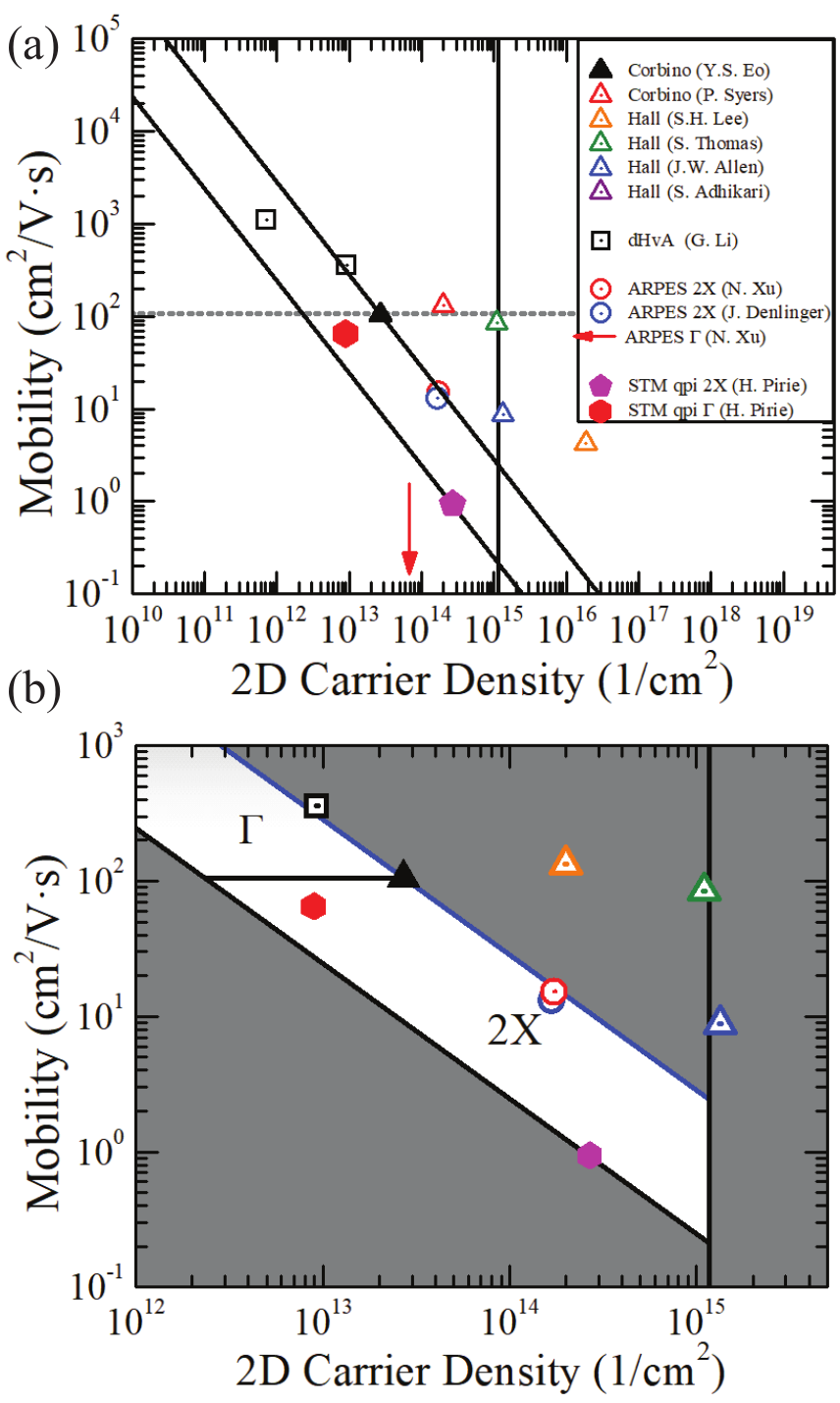}
\caption[Comparison of surface transport parameters of SmB$_{6}$ on the (001) surface.]{Comparison of surface transport parameters of SmB$_{6}$ on the (001) surface. (a) Comparison over a large transport range. (b) Comparison near the constrained parameter space. The triangular points are the transport reports. The square points are from the dHvA quantum oscillations. The circular points are from the ARPES $X$-pockets. The arrow is the carrier density estimation of the $\Gamma$-pocket from ARPES. The hexagonal points are from STM QPI reported pockets.}
\label{Fig:ComparisonParameterSpace}
\end{center}
\end{figure}

With the conversion relations that we have reviewed in the previous subsections, in Fig.~(\ref{Fig:ComparisonParameterSpace}) we plot the estimated mobility and carrier density of the reported transport, ARPES, dHvA, and STM QPI studies on the (001) surface of SmB$_{6}$. Because the figure is too crowded to plot all of the transport reports, we also provide a table of the transport reports, as shown in Table~\ref{Tab:TransportTable}, and the table of ARPES and STM QPI reports, as shown in Table~\ref{Tab:ARPESTable}, that also includes the data points that are omitted in the graph.

\begin{table*}[t]
\centering 
\begin{tabular}{c c c c c} 
\hline\hline 
Report & Geometry& R/square ($\Omega$)&$n_{2D}$ (cm$^{-2}$)& $\mu_{2D}$(cm$^{2}$/V$\cdot$sec)\\ [1ex]
\hline 
Y.S. Eo $et$ $al$. (This work)  & Corbino & 2.2$\times$10$^{3}$ & 2.71$\times$10$^{13}$ & 104.5\\ 
S. Wolgast $et$ $al$. (2013)\cite{Wolgast2} & Corbino & 5.1$\times$10$^{2}$&  2.0$\times$10$^{14}$ & 61\\
P. Syers $et$ $al$.(2015)\cite{PSyersCorbino} & Corbino & 2.4$\times$10$^{2}$ &2.0$\times$10$^{14}$ & 133\\
S. Thomas $et$ $al$. (2013)\cite{SThomasWAL} & Hall*  & 9.2$\times$10$^{-4}$&1.0$\times$10$^{18}$ & 6.8$\times$10$^3$\\
S.H. Lee $et$ $al$. (2013)\cite{SHLeeThinHall} & Thin Film Hall& 87 &1.86$\times$10$^{16}$ & 4.27\\
S. Adhikari $et$ $al$. (2015)\cite{adhikariHall} & Hall & 0.35& 2.0$\times$10$^{18}$ & 9\\
J.W. Allen $et$ $al$.(1979)\cite{JWAllenHall1979} & Hall* &5.3$\times$10$^{2}$& 1.34$\times$10$^{15}$ & 8.82\\
N.  J.  Laurita $et$ $al$.(2016)\cite{NLaurita} & ac Optical & $\geq$10$^3$& N/A & N/A\\[1ex] 
\hline 
\end{tabular}
\caption[Summary of the transport parameters from transport experiments.]{Summary of the transport parameters from transport experiments. *The carrier density and mobility are estimated naively by the Hall coefficient, thickness, and resistivity without the exact geometry information}
\label{Tab:TransportTable}
\end{table*}

\begin{table*}[t]
\centering 
\begin{tabular}{c c c c } 
\hline\hline 
Report & Pocket& $n_{2D}$ (cm$^{-2}$)& $\mu_{2D}$(cm$^{2}$/V$\cdot$sec)\\ [1ex]
\hline 
J. Denlinger $et$ $al$. (2013)\cite{Denlinger2013TempARPES,Denlinger2016consistency} & $2X$ & 1.67$\times$10$^{14}$ & 13\\ 
N. Xu $et$ $al$. (2013)\cite{NXuARPES2013} & $2X$ & 1.74$\times$10$^{14}$ & 15.2\\
N. Xu $et$ $al$.(2013)\cite{NXuARPES2013} & $\Gamma$ & 8.61$\times$10$^{12}$ & N/A\\
J. Jiang $et$ $al$. (2013)\cite{Jiang2ARPES2013}& $2X$  & 1.92$\times$10$^{14}$ & N/A\\
J. Jiang $et$ $al$. (2013)\cite{Jiang2ARPES2013} & $\Gamma$ &6.45$\times$10$^{12}$ & N/A\\
M. Neupane $et$ $al$. (2013)\cite{NeupaneARPES2013} & $2X$ & 3.01$\times$10$^{14}$ & N/A\\
M. Neupane $et$ $al$.(2013)\cite{NeupaneARPES2013} & $\Gamma$  & 9.39$\times$10$^{12}$ & N/A\\
P. Thunstrom $et$ $al$.(2019)\cite{thunstrom2019topology} & $\Gamma$  & 4.8$\times$10$^{13}$ & N/A\\
H. Pirie $et$ $al$.(2019)\cite{HarvardSTM, HarvardSTM_private} & $2X$  & 2.69$\times$10$^{14}$ & 0.94\\[1ex]
H. Pirie $et$ $al$.(2019)\cite{HarvardSTM, HarvardSTM_private} & $\Gamma$  & 8.94$\times$10$^{12}$ & 64.7\\[1ex]
\hline 
\end{tabular}
\caption[The transport parameters estimated from ARPES.]{The transport parameters estimated from ARPES and STM QPI.}
\label{Tab:ARPESTable}
\end{table*}

We explain some of the highlights of Fig.~(\ref{Fig:ComparisonParameterSpace})~(a) and tables Table~\ref{Tab:TransportTable} and Table~\ref{Tab:ARPESTable}. The transport reports, including our Corbino magnetotransport, are shown with triangles in Fig.~(\ref{Fig:ComparisonParameterSpace}). The early Hall effect data shows very high surface carrier density, with even exceeding the maximum carrier density associated with the BZ area. The very large carrier density tendency of these data points is consistent with our old estimation from the Hall geometry measurement. The ac optical conductivity study did not observe the existence of the surface states, but provide a minimum sheet conductivity based on their noise level\cite{NLaurita}. The dHvA quantum oscillations by G. Li $et$ $al$.\cite{GLiDhVA} report two Fermi pockets that originate from the (001) surface, plotted in hollow-square symbols in Fig.~(\ref{Fig:ComparisonParameterSpace}). The dHvA results show the highest mobility and lowest carrier density in the transport parameter space. Next, the ARPES reports of the $X$-pockets are plotted in circular points in Fig.~(\ref{Fig:ComparisonParameterSpace}). All of the ARPES studies that agree with the existence of the surface states see a very large $X$-pocket. Among these, J. Denlinger $et$ $al$.\cite{Denlinger2013TempARPES} and N. Xu $et$ $al$.\cite{NXuARPES2013} report the Im($\Sigma$), and therefore the mobility of the $X$-pocket can be estimated. Early ARPES studies report the $\Gamma$-pocket\cite{NXuARPES2013, NeupaneARPES2013, Jiang2ARPES2013} that has a small size. Recently, a suggestion that those states are trivial Rashba states has been made\cite{hlawenka2018samarium}. Instead, a larger $\Gamma$-pocket, previously interpreted as a Umklapp state formed folding of a $X$ pocket with respoect to the (2$\times$1) reconstruction surface Brillouin zone boundaries, has been proposed\cite{ohtsubo2019non,thunstrom2019topology}. Unfortunately, both reportes lack an estimation of Im($\Sigma$). Since we can only estimate the surface carrier density from the area of the pocket, we show one of the conversions from N. Xu $et$ $al.$\cite{NXuARPES2013} as an (orange) arrow pointing to the corresponding carrier density. Lastly, the STM QPI study report both the $X$- and $\Gamma$-pocket\cite{HarvardSTM}. Together with the scattering time that was estimated by the author\cite{HarvardSTM_private}, we report the estimated carrier density and mobility, as shown with hexagons in Fig.~(\ref{Fig:ComparisonParameterSpace}). 

\subsection{Estimation of Mobility of ARPES Fermi Pockets Using Transport}

In this subsection, by comparing our effective transport parameters from the Corbino disk magnetotransport with the ARPES reports that report a small $\Gamma$ pocket\cite{NXuARPES2013, NeupaneARPES2013, Jiang2ARPES2013}, we can further estimate what the mobility value of each pocket should be. Here we review two equations, Eq.~(\ref{Eq:totalcondCh3}) and Eq.~(\ref{Eq:totalmagncondCh3}).
\begin{equation}
\sigma_{t} = \sigma_{\Gamma} + \sigma_{2X},
\end{equation}
Furthermore, when a magnetic field is applied, we assume the carrier density and mobility we find from experiment is an effective parameter ($n_{\mathrm{eff}}$ and $\mu_{\mathrm{eff}}$) from the two pockets, i.e., $\sigma_{t}=n_{\mathrm{eff}}e\mu_{\mathrm{eff}}$. The magnetoconductivity is then: 
\begin{equation}
 \frac{n_{\mathrm{eff}}e\mu_{\mathrm{eff}}}{1+\mu_{\mathrm{eff}}^2B^2\cos^2\theta}
 =\frac{n_{\Gamma}e\mu_{\Gamma}}{1+\mu_{\Gamma}^2B^2\cos^2\theta}
 +\frac{n_{2X}e\mu_{2X}}{1+\mu_{2X}^2B^2\cos^2\theta}.
\end{equation}
Here, the known parameters are $n_{\mathrm{eff}}$ and $\mu_{\mathrm{eff}}$, where we use our magnetotransport results as well as $n_{\Gamma}$ and $n_{2X}$ from the ARPES reports in Table~(\ref{Tab:ARPESTable}). Using the two equations above, we can solve for $\mu_{\Gamma}$ and $\mu_{2X}$. We report the results in Table~(\ref{Tab:ARPESEstimate}). Our results predict that $\mu_{\Gamma}$ is about $\sim$150 cm$^2$/V$\cdot$s. A smaller $\Gamma$-pocket area requires a larger $\mu_{\Gamma}$ to be consistent with our magnetotransport report. 
 
\begin{table}[h]
\centering 
\begin{tabular}{c c c c } 
\hline\hline 
Report & $\mu_{\Gamma}$ (cm$^{2}$/V$\cdot$sec)& $\mu_{2X}$(cm$^{2}$/V$\cdot$sec)\\ [1ex]
\hline 
N. Xu $et$ $al$. (2014)\cite{NXuARPES2013} & 168.36 & 9.11\\
J. Jiang $et$ $al$. (2013)\cite{Jiang2ARPES2013} & 148.73 & 4.78\\
M. Neupane $et$ $al$.(2013)\cite{NeupaneARPES2013} & 152.96  & 8.71\\[1ex] 
\hline 
\end{tabular}
\caption[Estimation of mobility of the Fermi pockets ARPES after comparing with the Corbino magnetotransport.]{Estimation of mobility of the Fermi pockets ARPES after comparing with the Corbino magnetotransport.}
\label{Tab:ARPESEstimate}
\end{table}

\subsection{Consistency with Corbino Magnetotransport and Discussion}
	
In this subsection, we discuss how consistent the other previous studies are with our study of Corbino magnetotransport. We can say that they are in full agreement when the transport parameters ($n_{2D}$ and $\mu_{2D}$) of each pocket ($\Gamma$ and $2X$) from other studies are in the parameter space that we constructed in Fig.~(\ref{Fig:CorneredParameterSpace}). For convenience, we plot Fig.~(\ref{Fig:CorneredParameterSpace}) and Fig.~(\ref{Fig:ComparisonParameterSpace})~(a) together in  Fig.~(\ref{Fig:ComparisonParameterSpace})~(b). 

None of the previous studies are in full agreement with our transport study either because at least one of the pockets are lying exterior to the parameter space or full information is lacking (e.g., no mobility report). Although there is no full agreement, ARPES report by N. Xu $et.$ $al.$\cite{NXuARPES2013} are in most agreement among the ARPES studies. The 2$X$ pocket is within our constructed parameter space and the carrier density of $\Gamma$ is consistent with our expected range. J. Denlinger $et.$ $al.$\cite{Denlinger2013TempARPES} is in good agreement with our 2X parameter space, but the report of $\Gamma$ pocket is missing. The recent large $\Gamma$ pocket suggestion\cite{thunstrom2019topology, ohtsubo2019non} does not agree with our parameter space construction. In this case, the carrier density is close to the effective carrier density ($n_{eff}$) from the bare magnetotransport fitting. This requires an absence of the 2X pockets.  
The carrier density ($n_{2D}$) is the more important transport parameter than the mobility that needs agreement since it is a measure of the intrinsic property (electronic structure), whereas mobility can change based on the level of disorder of the sample and surface quality. Note in ARPES and STM experiments, the samples are cleaved and measured under vacuum. In contrast in transport, the samples are exposed in ambient conditions. We therefore believe a mobility difference by a few factors and not orders of magnitude difference can have potential agreement. In that sense, the STM QPI study are in most agreement except for its consistently lower mobilities. From STM QPI, the tendency that the $\Gamma$ pocket has a higher mobility than the $2X$ pocket is consistent with our transport analysis. One might attribute the consistently lower mobilities are from the sample and surface quality difference. Interestingly, the authors from the same group provide a scenario that may be an alternative explanation for this disagreement. The authors explain the difference in ARPES and STM QPI, most notably the difference in effective mass can be explained by band bending driven by polar reconstruction surfaces\cite{matt2018consistency}. The authors note that the notable difference between STM QPI and ARPES are the length scales they are probing. The largest ordered domains measured by STM are on the order of tens of nanometers, whereas ARPES measures a spatial average of tens of microns consisting of many domains formed by reconstruction surfaces that may be either polar or non-polar\cite{matt2018consistency}. The dominant effect on polar surfaces is to shift the Dirac point of the surface towards the bulk valance band and therefore results in an enhanced Fermi velocity. The study shows that a smaller effective mass can be seen by a spatial averaging including these polar reconstruction surfaces\cite{matt2018consistency}. A full consistency with our transport, that is also agreeing with the mobility of the $\Gamma$ pocket, might be possible after considering a further spatial averaging\cite{HarvardSTM_private}. 
The pockets reported by dHvA are more difficult to agree with our constructed parameter space. All of the pocket sizes are small. The large pocket size that is consistent with the $X$-pocket parameter space seems to be lacking. From the perspective of our transport parameter space, an extremely large magnetic field ($>$1000 T) is needed for this observation. Instead, the two pockets in dHvA are reported both closer to the $\Gamma$ pocket parameter space. Also, the two pockets have higher mobilities than the effective mobility we find from magnetotransport (see Appendix A for reasoning). We note that the mobility estimated from the dHvA oscillations is the quantum mobility ($\mu_Q$) and not the transport mobility ($\mu$). Still, this cannot explain the disagreement since typically, $\mu_Q$ and $\mu$ only differ slightly. Finally, we briefly mention the Cambridge dHvA results, although our study does not aim to re-interpret the bulk origin claim to the surface. There has been a study that re-considers the quantum oscillations from the Cambridge group by Tan $et$ $al.$\cite{TanDhVa} as having a surface origin instead of a bulk origin\cite{Denlinger2016consistency}. The corresponding Fermi surface area of the high frequencies is consistent with the X-pocket observed by ARPES. When comparing the Cambridge quantum oscillations with our transport results, the corresponding carrier density is also consistent with the 2X parameter space that we constructed. However, the associated mobility for the onset of those high-frequency quantum oscillations is far too high to be consistent with our transport parameter space and the ARPES reports.
 
\section{Conclusion} 
 
In conclusion, we have studied the surface transport of SmB$_{6}$ using high-field magnetotransport at low temperatures. We have not been able to observe SdH quantum oscillations in our studies, and therefore we were not able to see signatures of multiple Fermi pockets. To properly characterize the surface without SdH oscillations, we have discussed the importance of employing the proper transport geometry and surface preparation in surface transport studies. From our estimated carrier density and mobility, we were able to constrain the possible carrier density and mobility values of each Fermi pocket (each channel).

\begin{acknowledgments}
We thank H. Pirie for providing us the scattering time estimations from recent STM QPI work and the helpful discussions. We also thank K. Sun, L. Li, and J. W. Allen for their insightful comments. Portion of this work performed at the National High Magnetic Field Laboratory (NHMFL), which is supported by the National Science Foundation Cooperative Agreement Nos. DMR-1157490 and DMR-1644779 and the State of Florida, and by the DOE. We thank Jan Jaroszynski for assistance during the high magnetic field experiments. Device fabrication was performed in part at the Lurie Nanofabrication Facility. The authors acknowledge the University of Michigan College of Engineering for financial support and the Michigan Center for Materials Characterization for use of the instruments and staff assistance.
\end{acknowledgments} 

\appendix
\section{\label{Appendix:ParaSpace}Details on the Parameter Space Construction }

In this appendix, we discuss in more detail some considerations in constructing the parameter space of Fig.~(\ref{Fig:CorneredParameterSpace}). 

\subsection{Mobilities of the Two Pockets} 
For the reader, it may not be obvious why one of the channels ($\Gamma$) requires a higher mobility than the effective mobility (the mobility we find from the surface Corbino magnetotransport experiment), while the other channel ($2X$) requires a lower mobility than the effective mobility. 

First, suppose the mobilities from both pockets ($\mu_{\Gamma}$ and $\mu_{2X}$) are larger than the effective mobility $\mu_{\mathrm{eff}}$, i.e., $\mu_{\Gamma} > \mu_{\mathrm{eff}}$ and $\mu_{2X} > \mu_{\mathrm{eff}}$. We can express Eq.~(\ref{Eq:EffectMobility}) as:
\begin{equation}
\mu_{\mathrm{eff}}^2 =(\mu_{\Gamma}^2-\mu_{2X}^2)(\frac{\sigma_{\Gamma}}{\sigma_{t}})+\mu_{2X}^2 = \frac{\sigma_{\Gamma} \mu_{\Gamma}^2 + \sigma_{2X} \mu_{2X}^2 }{\sigma_{t}},
\label{Eq:EffectMobilityAppendix}
\end{equation}

We first consider the condition of $\mu_{\Gamma} > \mu_{\mathrm{eff}}$. This requires:
\begin{equation}
\mu_{\Gamma}^2 > \frac{\sigma_{\Gamma} \mu_{\Gamma}^2 + \sigma_{2X} \mu_{2X}^2 }{\sigma_{t}}.
\label{Eq:EffectMobilityGamma1}
\end{equation}
Eq.~(\ref{Eq:EffectMobilityGamma1}) can be rearranged as 
\begin{equation}
(\sigma_t - \sigma_{\Gamma})~\mu_{\Gamma}^2 > \sigma_{2X} ~\mu_{2X}^2.
\label{Eq:EffectMobilityGamma2}
\end{equation}
Since $\sigma_{t} = \sigma_{\Gamma} + \sigma_{2X}$, we can write Eq.~(\ref{Eq:EffectMobilityGamma2}) as:
\begin{equation}
\sigma_{2X}~\mu_{\Gamma}^2 > \sigma_{2X}~ \mu_{2X}^2.
\label{Eq:EffectMobilityGamma3}
\end{equation}

Since the conductivity and mobility are always positive, we can cancel the conductivity and take the square root of mobility without changing the inequality sign. This reduces to:
\begin{equation}
\mu_{\Gamma} > \mu_{2X}.
\label{Eq:EffectMobilityGamma5}
\end{equation}

Next we consider $\mu_{2X} > \mu_{\mathrm{eff}}$. We can again use Eq.~(\ref{Eq:EffectMobilityAppendix}):
\begin{equation}
\mu_{2X}^2 > \frac{\sigma_{\Gamma} \mu_{\Gamma}^2 + \sigma_{2X} \mu_{2X}^2 }{\sigma_{t}}.
\label{Eq:EffectMobility2X1}
\end{equation}
Then, similar to the steps Eq.~(\ref{Eq:EffectMobilityGamma1}) - Eq.~(\ref{Eq:EffectMobilityGamma5}), Eq.~(\ref{Eq:EffectMobility2X1}) can be expressed as:

\begin{equation}
\mu_{2X} > \mu_{\Gamma}.
\label{Eq:EffectMobility2X3}
\end{equation} 

Eq.~(\ref{Eq:EffectMobility2X3}) is in contradiction to Eq.~(\ref{Eq:EffectMobilityGamma5}), so $\mu_{\Gamma} > \mu_{\mathrm{eff}}$ and $\mu_{2X} > \mu_{\mathrm{eff}}$ cannot be satisfied at the same time. 

Next, we consider when at least one of the mobilities is the same as the effective mobility, $\mu_{\Gamma} = \mu_{\mathrm{eff}}$ or (and) $\mu_{2X} = \mu_{\mathrm{eff}}$. We consider $\mu_{\Gamma} = \mu_{\mathrm{eff}}$ case first. From, Eq.~(\ref{Eq:EffectMobilityAppendix}), this condition requires:

\begin{equation}
\mu_{\Gamma}^2 = \frac{\sigma_{\Gamma} \mu_{\Gamma}^2 + \sigma_{2X} \mu_{2X}^2 }{\sigma_{t}}.
\label{Eq:EffectMobilityGamma6}
\end{equation}

Then, similar to the  calculations above, Eq.~(\ref{Eq:EffectMobilityGamma1}) results in:
\begin{equation}
\mu_{\Gamma} = \mu_{2X}.
\label{Eq:EffectMobilityGamma7}
\end{equation}

Similarly, from the condition $\mu_{2X} = \mu_{\mathrm{eff}}$, we get an identical result. Because Eq.~(\ref{Eq:totalcondCh3}), $\sigma_{t} = \sigma_{\Gamma} + \sigma_{2X}$ must also satisfy $n_{\Gamma} = n_{2X}$. This result is meaningless in that we have just split one pocket into two pockets with identical properties. 

Therefore, in the presence of two channels, one of the mobilities has to be larger than the effective mobility ($\mu_{\Gamma} > \mu_{\mathrm{eff}}$), and the other has to be smaller ($\mu_{2X} < \mu_{\mathrm{eff}}$):

\begin{equation}
\mu_{2X} < \mu_{\mathrm{eff}} < \mu_{\Gamma}.
\label{Eq:EffectMobilityResult}
\end{equation}

\subsection{Carrier Densities of the Two Pockets} 

Next, we consider the carrier densities of each pocket in the parameter space.  For the reader, it may also not be clear why one of the channels ($\Gamma$) requires a higher carrier density than the effective carrier density (the carrier density we find from the Corbino magnetotransport), while the other channel ($2X$) has to occupy the lower right part of the parameter space region.  

Notice that because the total conductivity is the sum of the conductivities of each channel, the conductivity of each channel must be smaller than the total conductivity, $\sigma_{t} > \sigma _{\Gamma}$ and $\sigma_{t} > \sigma _{2X}$.

We first consider the condition of  $\sigma_{t} > \sigma _{\Gamma}$. This requires:

\begin{equation}
n_{\mathrm{eff}}\mu_{\mathrm{eff}} > n_{\Gamma}\mu_{\Gamma}.
\label{Eq:CarrierDensityGamma1}
\end{equation}

Then, we can express this as:

\begin{equation}
\frac{\mu_{\mathrm{eff}}}{\mu_{\Gamma}} > \frac{n_{\Gamma}}{n_{\mathrm{eff}}}.
\label{Eq:CarrierDensityGamma2}
\end{equation}

Since $\mu_{\mathrm{eff}}/\mu_{\Gamma}$ must be smaller than 1 from what we found from the previous section, $n_{\Gamma}/\mu_{\mathrm{eff}}$ must also be smaller than 1. Then we find:

\begin{equation}
n_{\mathrm{eff}} > n_{\Gamma}.
\label{Eq:CarrierDensityGamma3}
\end{equation}

Next, for the carrier density for the $2X$ channel, since $\sigma_{2X} > (e^2/h)$, and $\mu_{\mathrm{eff}}>\mu_{2X}$, we have the following inequality:

\begin{equation}
n_{2X} > \frac{e}{h}\frac{1}{\mu_{2X}}>\frac{e}{h}\frac{1}{\mu_{\mathrm{eff}}}.
\label{Eq:CarrierDensity2X}
\end{equation}

Therefore, $n_{2X} > 2.3\times 10^{12}$ cm$^{-2}$. Together with $\mu_{\mathrm{eff}} > \mu_{2X}$, this corresponds to the lower right region of Fig.~(\ref{Fig:CorneredParameterSpace}).
\section{\label{Appendix:methods} Sample Preparation}
In this section, we discuss how the samples were prepared. In this work, we tried SmB$_6$ crystals grown by both the Al flux method and the floating zone method from different crystal growth facillities. Samples that were grown by Al flux were grown by University of California - Irvine (UCI) by Z. Fisk's group, Gwangju Institute of Science and Technology, S. Korea by B. Cho's group, and University of Maryland by J. Paglione's group. The floating zone samples were grown by University of Warwick, UK, by G. Balakrishnan's group. 

Some of the early samples that were used in Ref. ~\cite{Wolgast2} were re-used in this work. In Ref. ~\cite{Wolgast2}, these samples were polished with a SiC abrasive pads with P4000 as the finest polishing step. For the scratch experiment we used the Fisk011 sample. FiskGate1 sample was re-polished as described in the following. 

For the samples that were used for magnetotransport studies at the National High Magnetic Field Lab (cell12), we prepared the samples with a different polishing method. Freshly grown samples were polished by a lapping machine (Logitech Pm5 lapper) starting with a 3 $\mu$m Al$_2$O$_3$ powder slurry mixed with deionized water for shaping the material. Then, finer polishing of 1 $\mu$m and then 0.3 $\mu$m were used for optical polishing. The fine polishing step was also thinned with length scales larger than the previous grit size. We later tested samples with as-grown surfaces as shown in Fig.~\ref{Fig:AsGrown} in our own lab measured with the PPMS Dynacool 14T rotator option. These samples were not polished. 
Polished surfaces or as-grown surfaces were etched with diluted HCl before Corbino disk patterns were fabricated. The Corbino disks were patterned using standard photolithography. The patterns consist of 20-30 $\AA$/1500 $\AA$ Ti/Au Contacts by ebeam evaporation. 1mil Al and Au wires were used by wire bonding. 
\section{\label{Appendix:Sample Variation}Sample Variation }
\begin{figure}[t]
\begin{center}
\includegraphics[scale=0.8]{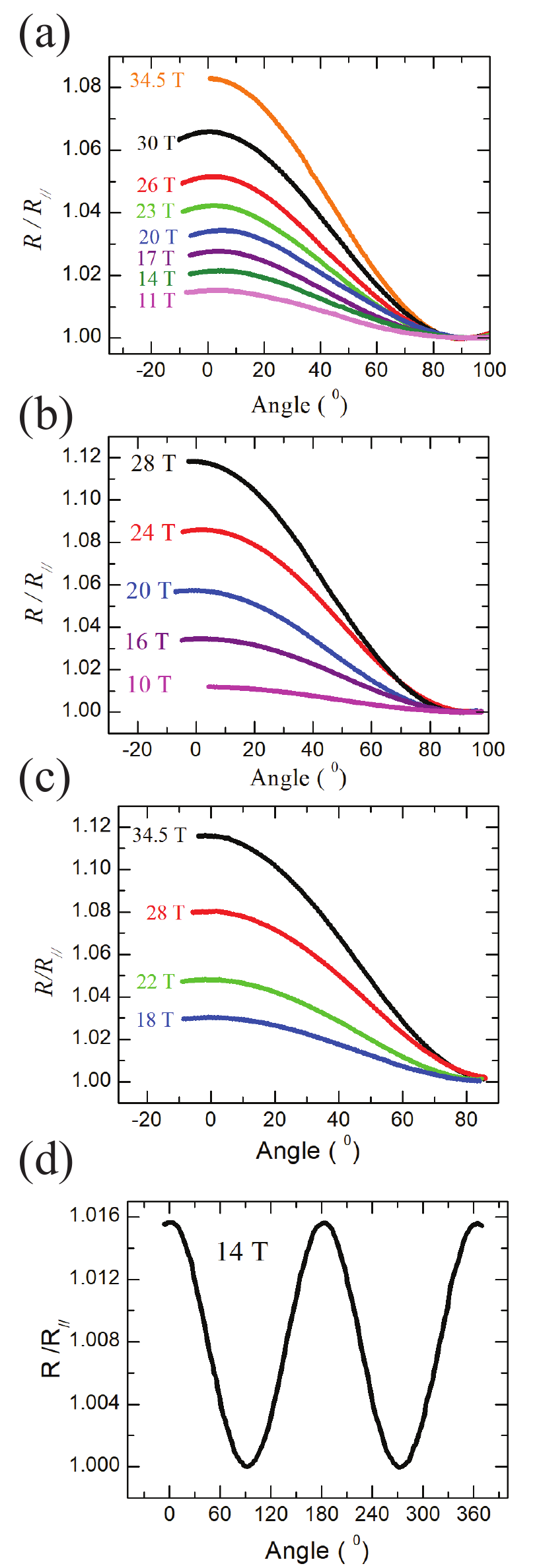}
\caption[Angle Dependent Corbino Resistance at Fixed Magnetic Fields.]{Angle Dependent Corbino Resistance at Fixed Magnetic Fields. (a) A Freshly prepared sample from Z. Fisk's group (Sample: FiskGate2) (b) Sample grown by floating zone. Corbino disk prepared on the (011) polished surface. (Sample: Warwick011) (c) Sample grown by floating zone. Corbino disk prepared on the (111) polished surface. (Sample Warwick111) (d) Sample prepared on a clean as grown surface. Note: (a)-(c) were measured at 0.35K and (d) was measured at 1.8K in PPMS Dynacool rotator option.) }
\label{Fig:SampleDepMR}
\end{center}
\end{figure}

In this appendix, we discuss the sample dependence of the 2D carrier density and mobility. In the main text, we only shown the magnetotransport result of only one of the samples (FiskGate1 sample). Here, we discuss the variation results from multiple samples we have tested. 
In Fig.~(\ref{Fig:SampleDepMR}), we show the angle dependence of Corbino resistance at a fixed magnetic field. We find that all samples show a carrier density and mobility in the range of 90-120 (cm$^2$/V-sec) and 2-3 $\times$ 10$^{13}$ (1/cm$^2$), which is consistent with the results of what was discussed in the main text.  

\bibliography{ref}

\end{document}